\documentclass[preprint]{aastex}
\usepackage[usenames]{color}

\begin{document}

\shortauthors{Welty et al.}
\shorttitle{Magellanic Clouds Thermal Pressures}             


\title{Thermal Pressures in the Interstellar Medium of the Magellanic Clouds\footnotemark}
\footnotetext{Based on observations made with the NASA/ESA {\it Hubble Space Telescope}, obtained at the Space Telescope Science Institute, which is operated by the Association of Universities for Research in Astronomy, Inc., under NASA contract NAS 5-26555. These observations are associated with programs 8145, 9383, 9757, and 12978.
Based on observations made with ESO Telescopes at the La Silla Paranal Observatory, under program IDs 062.I-0841, 064.I-0475, 066.C-0365, and 072.C-0682, and on data obtained from the ESO Science Archive Facility by user DWELTY}

\author{Daniel E. Welty\altaffilmark{1}, James T. Lauroesch\altaffilmark{2}, Tony Wong\altaffilmark{3}, Donald G. York\altaffilmark{1,4}}

\altaffiltext{1}{University of Chicago, Department of Astronomy and Astrophysics, 5640 S. Ellis Ave., Chicago, IL 60637; dwelty@oddjob.uchicago.edu}
\altaffiltext{2}{University of Louisville, Department of Physics and Astronomy, Louisville, KY 40292}
\altaffiltext{3}{University of Illinois at Urbana/Champaign, Department of Astronomy, 1002 W. Green St., Urbana, IL 61801}
\altaffiltext{4}{also, Enrico Fermi Institute}

\begin{abstract}

We discuss the thermal pressures ($n_{\rm H}T$) in predominantly cold, neutral interstellar gas in the Magellanic Clouds, derived from analyses of the fine-structure excitation of neutral carbon, as seen in high-resolution {\it HST}/STIS spectra of seven diverse sight lines in the LMC and SMC.  
Detailed fits to the line profiles of the absorption from \ion{C}{1}, \ion{C}{1}*, and \ion{C}{1}** yield consistent column densities for the 3--6 \ion{C}{1} multiplets detected in each sight line.  
In the LMC and SMC, $N$(\ion{C}{1}$_{\rm tot}$) is consistent with Galactic trends versus $N$(\ion{Na}{1}) and $N$(CH), but is slightly lower versus $N$(\ion{K}{1}) and $N$(H$_2$).
As for $N$(\ion{Na}{1}) and $N$(\ion{K}{1}), $N$(\ion{C}{1}$_{\rm tot}$) is generally significantly lower, for a given $N$(H$_{\rm tot}$), in the LMC and (especially) in the SMC, compared to the local Galactic relationship.
For the LMC and SMC components with well determined column densities for \ion{C}{1}, \ion{C}{1}*, and \ion{C}{1}**, the derived thermal pressures are typically factors of a few higher than the values found for most cold, neutral clouds in the Galactic ISM.
Such differences are consistent with the predictions of models for clouds in systems (like the LMC and SMC) that are characterized by lower metallicities, lower dust-to-gas ratios, and enhanced radiation fields -- where higher pressures are required for stable cold, neutral clouds.
The pressures may be further enhanced by energetic activity (e.g., due to stellar winds, star formation, and/or supernova remnants) in several of the regions probed by these sight lines.
Comparisons are made with the \ion{C}{1} observed in some quasar absorption-line systems.

\end{abstract}


\section{INTRODUCTION}
\label{sec-intro}

The total interstellar pressure required for hydrostatic equilibrium in the Galactic disk, of order $p_{\rm tot}/k$ $\sim$ 2.8 $\times$ 10$^4$ cm$^{-3}$K, includes contributions from magnetic fields, cosmic rays, and both thermal and turbulent motions of the gas (e.g., Boulares \& Cox 1990).
Models of interstellar clouds that include all significant heating and cooling processes (e.g., Wolfire et al. 1995, 2003) predict two stable phases for predominantly neutral clouds -- a cold neutral phase (CNM), with $T$ $\la$ 200 K, and a warm neutral phase (WNM), with $T$ $\sim$ 10$^4$ K.
For Wolfire et al.'s (2003) ''standard model'' of the local Galactic interstellar medium (ISM), both of those phases can be present for thermal pressures $p/k$ = $n_{\rm H}T$ $\sim$ 2.0--4.8 $\times$ 10$^3$ cm$^{-3}$K.
While the thermal pressure thus is generally only a minor contributor to the total interstellar pressure, it nevertheless provides important information on the local physical conditions and processes in the ISM.
For example, the thermal pressure apparently is proportional to both the (larger) turbulent pressure (Joung et al. 2009) and the local heating and star-formation rates (Ostriker et al. 2010).

For relatively cold, neutral gas, the thermal pressure is often estimated via analysis of the excitation of neutral carbon, for which the populations of the two upper $^3P_1$ and $^3P_2$ fine-structure levels in the ground electronic state (denoted \ion{C}{1}* and \ion{C}{1}**, with excitation energies $E/k$ = 23.6 and 62.4 K, respectively) are determined principally by a balance between collisional excitation, collisional de-excitation, and radiative decay (Jenkins \& Shaya 1979).
A recent general survey of the \ion{C}{1} absorption and excitation in the Galactic ISM, based on high-resolution UV spectra of 89 stars obtained with the Space Telescope Imaging Spectrograph (STIS) on board the {\it Hubble Space Telescope} ({\it HST}), has indicated that most of the gas traced by \ion{C}{1} is characterized by relatively low thermal pressures, but that there is generally also a small fraction of gas (typically less than 10\%) at much higher pressures (Jenkins \& Tripp 2011; hereafter JT11; see also Jenkins \& Tripp 2001 and Burgh et al. 2010).
The dominant low-pressure component corresponding to the ''center of mass'' of the distribution of \ion{C}{1} populations in JT11's full data set has log($p/k$)$_{\rm low}$ $\sim$ 3.60 (cm$^{-3}$K), which corresponds to a local hydrogen density $n_{\rm H}$ = 50 cm$^{-3}$ for a representative $T$ = 80 K.

The Magellanic Clouds are characterized by lower metallicities [factors of about 2 (LMC) and 5 (SMC) below Solar values], dust-to-gas ratios correspondingly lower than those found in the local Galactic ISM (e.g., Welty et al. 2012; Roman-Duval et al. 2014), and generally stronger UV radiation fields (e.g., Lequeux 1989; Bernard et al. 2008; Sandstrom et al. 2010).
Those environmental differences are predicted to affect the structure, chemistry, and physical characteristics of interstellar clouds in the LMC and SMC (e.g., Maloney \& Black 1988; Wolfire et al. 1995, 2003; Johansson 1997; Pak et al. 1998).
In particular, photodissociation regions are expected to be more extensive, molecular cores (as traced by CO emission) are expected to be smaller, and somewhat higher thermal pressures should be needed to maintain stable cold, neutral clouds in such environments.
Some evidence for the predicted higher thermal pressures in diffuse atomic and molecular gas in the Magellanic Clouds may be gleaned from moderate-resolution UV spectra of \ion{C}{1} (Welty et al. 1999a; Koenigsberger et al. 2001; Andr\'{e} et al. 2004), from the elevated $J$=4,5 populations of H$_2$ (Tumlinson et al. 2002; Cartledge et al. 2005; Xue et al., in preparation), and from the rotational excitation of C$_2$ toward one SMC star (Welty et al. 2013).
The limited resolution of the previously reported \ion{C}{1} and H$_2$ spectra, however, generally allows only average values of the excitation to be estimated for those typically complex lines of sight; higher resolution spectra are needed to determine the characteristics of individual clouds in the LMC and SMC.
Higher densities have also been inferred for denser clouds more closely related to star formation (e.g., van Loon et al. 2010), consistent with theoretical expectations for lower metallicity systems (e.g., Krumholz et al. 2009).
And while [\ion{C}{1}] emission from the first excited fine-structure level ($^3P_1$ $\rightarrow$ $^3P_0$; at 609 $\mu$m) has been detected at several locations in the LMC and SMC (Stark et al. 1997; Bolatto et al. 2000a, 2000b), with $I_{\rm [C I]}$/$I_{\rm CO}$ comparable to or slightly greater than the values found for similar Galactic regions, the total abundance and excitation of \ion{C}{1} cannot be directly determined from those observations.

A more detailed understanding of the abundance and excitation of \ion{C}{1} in the Magellanic Clouds should also aid in understanding the \ion{C}{1} seen in some quasar absorption-line systems -- many of which also have sub-Solar metallicities (Songaila et al. 1994; Roth \& Bauer 1999; Quast et al. 2002; Srianand et al. 2005; Jorgenson et al. 2010; Carswell et al. 2011; Ledoux et al. 2015).
Detectable \ion{C}{1} may be a useful indicator of the presence of colder, denser gas with higher molecular content in those more distant systems (Papadopoulos et al. 2004; Srianand et al. 2005; Glover \& Clark 2016), and its excitation can provide both information on local physical conditions and constraints on the redshift dependence of the temperature of the Cosmic Microwave Background (CMB).

In this paper, we describe analyses of the \ion{C}{1} fine-structure excitation in the ISM of the Magellanic Clouds -- based on the \ion{C}{1} absorption seen in high-resolution STIS echelle spectra of seven stars in the LMC and SMC -- which indicate systematically higher thermal pressures there.
Section~\ref{sec-obs} describes the sight line sample, the optical and UV spectra obtained for the seven stars, and the analysis of the absorption-line profiles seen in those spectra.
Section~\ref{sec-res} presents the abundances and excitation of \ion{C}{1} derived from the UV spectra and the thermal pressures estimated from the \ion{C}{1} excitation.
Section~\ref{sec-disc} discusses the newly derived thermal pressures, in relation to cloud models and to the environments probed by the seven SMC and LMC sight lines, and makes comparisons with the \ion{C}{1} seen in more distant quasar absorption-line systems.
Section~\ref{sec-sum} gives a summary of our results and conclusions.
Several appendices present the full set of \ion{C}{1} profiles, the detailed component structures derived from our profile fits, and compilations of the Galactic column densities, Magellanic Clouds carbon abundances, and quasar absorber \ion{C}{1} column densities used in the main body of the paper.
A subsequent paper will focus on the elemental abundances and depletions found for five of these sight lines (Welty et al., in preparation).

\begin{deluxetable}{llrrrrrlr}
\tablecolumns{9}
\tabletypesize{\scriptsize}
\tablecaption{SMC and LMC Sight Lines \label{tab:los}}
\tablewidth{0pt}

\tablehead{
\multicolumn{1}{c}{Star}&
\multicolumn{1}{c}{Name}&
\multicolumn{2}{c}{RA (J2000) Dec}&
\multicolumn{1}{c}{$V$}&
\multicolumn{1}{c}{$(B-V)$}&
\multicolumn{1}{c}{$E(B-V)$}&
\multicolumn{1}{c}{Type}&
\multicolumn{1}{c}{Program}\\
\multicolumn{1}{c}{ }&
\multicolumn{1}{c}{ }&
\multicolumn{1}{c}{($^{h~m~s}$)}&
\multicolumn{1}{c}{($\arcdeg~\arcmin~\arcsec$)}&
\multicolumn{2}{c}{ }&
\multicolumn{1}{c}{tot/MC}&
\multicolumn{1}{c}{ }&
\multicolumn{1}{c}{ }}

\startdata
Sk 13       & AzV 18    & 00 47 12.2 & $-$73 06 33 & 12.44 &    0.03 & 0.20/0.16 & B2 Ia    &  9383 \\ 
Sk 18       & AzV 26    & 00 47 50.0 & $-$73 08 21 & 12.46 & $-$0.17 & 0.15/0.11 & O7 III   & 12978 \\ 
Sk 143      & AzV 456   & 01 10 55.8 & $-$72 42 56 & 12.83 &    0.10 & 0.36/0.33 & O9.7 Ib  &  9383 \\ 
Sk 155      & AzV 479   & 01 14 50.3 & $-$73 20 18 & 12.46 & $-$0.15 & 0.13/0.10 & O9 Ib    &  8145 \\ 
\hline
Sk$-$67~5   & HD 268605 & 04 50 18.9 & $-$67 39 38 & 11.34 & $-$0.12 & 0.14/0.11 & O9.7 Ib  &  9757 \\ 
Sk$-$68~73  & HD 269445 & 05 22 59.8 & $-$68 01 47 & 11.45 &    0.27 & 0.40/0.34 & Of/WN    & 12978 \\ 
Sk$-$70~115 & HD 270145 & 05 48 49.7 & $-$70 03 58 & 12.24 & $-$0.10 & 0.20/0.14 & O6.5 Iaf &  9757 \\ 
\enddata
\tablecomments{The first four stars are in the SMC; the last three stars are in the LMC.
Values for photometry and spectral type are from Welty et al. 2012, where references to the sources of those data are listed.}
\end{deluxetable}

\begin{deluxetable}{lcccccccll}
\tablecolumns{10}
\tabletypesize{\scriptsize}
\tablecaption{SMC and LMC Environments \label{tab:env}}
\tablewidth{0pt}

\tablehead{
\multicolumn{1}{c}{Star}&
\multicolumn{1}{c}{$v_{\rm rad}$}&
\multicolumn{1}{c}{Ref}&
\multicolumn{1}{c}{$v$(ISM)}&
\multicolumn{1}{c}{$v$(CO)}&
\multicolumn{1}{c}{Ref}&
\multicolumn{1}{c}{$v$(H II)}&
\multicolumn{1}{c}{Ref}&
\multicolumn{1}{c}{H II Name}&
\multicolumn{1}{c}{Comments}}

\startdata
Sk~13       & 138/147 & 1 & 148 & 158 & 1 & 125     & 1 & N19; DEM 31     & complex region\\
Sk~18       & 142/149 & 2 & 124 & 158 & 1 & 143     & 1 & N19; DEM 32     & complex region\\
Sk~143      & 167/... & 3 & 133 & ... &   & 140/170 & 1 & DEM 140         & isolated, faint nebula\\
Sk~155      & 173/158 & 2 & 160 & 179 & 2 & 174     & 1 & N84B; DEM 152   & no associated nebula \\
\hline
Sk$-$67~5   & 309/... & 4 & 288 & 287 & 1 & ...     &   & N3; DEM 7       & isolated, faint nebula\\
Sk$-$68~73  & 278/... & 5 & 296 & 296 & 3 & 301     & 2 & N44H; DEM 160   & complex region\\
Sk$-$70~115 & 337/278 & 6 & 220 & 238 & 3 & 242     & 3 & N180AB; DEM 323 & complex region? \\
\enddata
\tablecomments{Velocities (km~s$^{-1}$) are heliocentric. 
First stellar radial velocity is from the literature; second is from Welty \& Crowther, in preparation.
ISM velocity is for the strongest Na~I component in Table~\ref{tab:comps}.
Velocities for CO and H~II are for the nearest observed molecular cloud or H~II region (as listed).}
\tablerefs{$v_{\rm rad}$: 1 = Evans et al. 2004; 2 = Maurice et al. 1989; 3 = Evans \& Howarth 2008; 4 = Walborn et al. 2002; 5 = Feast et al. 1960; 6 = Neugent et al. 2012.
$v$(CO): 1 = Israel et al. 1993, 2003; 2 = Bolatto et al. 2003; 3 = Wong et al. 2011.
$v$(H~II): 1 = Le Coarer et al. 1993; 2 = Ch\'{e}riguene \& Monnet 1972; 3 = Georgelin et al. 1983.}
\end{deluxetable}

\section{OBSERVATIONS AND DATA ANALYSIS}
\label{sec-obs}

\subsection{Sight lines}
\label{sec-los}

Basic data for the seven sight lines included in this study are given in Table~\ref{tab:los}; some additional information on nearby (and possibly associated) CO emission and \ion{H}{2} regions is given in Table~\ref{tab:env}.
All seven sight lines have been observed with both {\it HST}/STIS and {\it FUSE} in the UV -- yielding column densities for both H and H$_2$ (Tumlinson et al. 2002; Cartledge et al. 2005; Welty et al. 2012; Xue et al., in preparation) and for various atomic and molecular species (Friedman et al. 2000; Welty et al. 2001, 2004, and in preparation; Andr\'{e} et al. 2004; Sofia et al. 2006; Tchernyshyov et al. 2015).
The total neutral hydrogen column densities $N$(H$_{\rm tot}$) = $N$(H) + 2$N$(H$_2$) range from about 1.1--11.5 $\times$ 10$^{21}$ cm$^{-2}$, with molecular fractions $f$(H$_2$) = 2$N$(H$_2$)/$N$(H$_{\rm tot}$) from 0.01--0.63.
All seven have also been observed with ground-based spectrographs, for investigations of various constituents of the ISM in the SMC and LMC (e.g., Caulet \& Newell 1996; Cox et al. 2006, 2007; Welty et al. 2006, 2013; Howk et al. 2012; Welty \& Crowther 2010 and in preparation).\footnotemark
\footnotetext{Some information about these and other SMC and LMC sight lines, with plots of the optical spectra, may be found at http://astro.uchicago.edu/$\sim$dwelty/mcoptuv.html.
Optical/IR images of the regions around each star and links to the {\it FUSE} spectra may be found at the {\it FUSE} Magellanic Clouds Legacy Project website (https://archive.stsci.edu/prepds/fuse\_mc/; Blair et al. 2009; see also Danforth et al. 2002).}
While most of the stars appear to be located in or near \ion{H}{2} regions and/or significant amounts of molecular material (e.g., Henize 1956; Davies et al. 1976; Israel et al. 1993, 2003; Bica et al. 2008; Wong et al. 2011), the sight lines probe a number of different regions in the SMC and LMC, and the interstellar material observed along them exhibits a range of properties:
\begin{itemize}
\item{Both Sk~13 and Sk~18 are located in the N19 complex of \ion{H}{2} regions, supernova remnants, and molecular gas in the southwest part of the main SMC ''bar'', near DEM~31 and DEM~32, respectively (Rubio et al. 1993; Rosado et al. 1994; Dickel et al. 2001; Bot et al. 2010).
The sight line to Sk~13 exhibits ''typical'' SMC extinction, with no 2175~\AA\ bump and a very steep far-UV rise (Gordon et al. 2003).
Sofia et al. (2006) have discussed the abundances of several species detected in the STIS spectra of Sk~13; the overall Ti depletion\footnotemark\ toward Sk~18 is relatively mild, despite the fairly high molecular fraction (Welty \& Crowther 2010).}
\footnotetext{In low-metallicity systems, the depletions are determined relative to the overall metallicity of the system.
For the SMC and LMC, the metallicities have been estimated from the abundances of various elements measured in relatively young stars and nebulae.
For quasar absorption-line systems, the metallicities are estimated from the gas-phase abundances of typically mildly depleted elements (e.g., Zn, or perhaps O, Si, and/or S), relative to H$_{\rm tot}$.}
\item{Sk~143 is located between the northeastern end of the SMC ''bar'' and the SMC ''wing'' region, and appears to be associated with the diffuse \ion{H}{2} region DEM~140.
This sight line is unusual in the SMC, with the UV extinction curve, the dust-to-gas ratio, and the relative abundances of a number of atomic and molecular species more similar to those found in the local Galactic ISM (Gordon et al. 2003; Welty et al. 2006, 2012; Cox et al. 2007).
To this point, it has both the only detections of \ion{Li}{1}, CN, C$_2$, and C$_3$ absorption and the most severe depletion of Ti known in the SMC; the ratio of the local hydrogen density to the strength of the UV radiation field ($n_{\rm H}$/$I_{\rm UV}$) appears to be higher than for other sight lines in the SMC (Welty et al. 2006, 2013; Welty \& Crowther 2010; Howk et al. 2012; see also Sofia et al. 2006).
CO is detected in absorption in the STIS spectrum, but has not been detected in emission toward Sk~143 (Rubio et al. 1993b).}
\item{Sk~155 is the only blue supergiant in the NGC~460b stellar association, located toward the N84 complex of \ion{H}{2} regions and molecular gas in the near ''wing'' region of the SMC (Bolatto et al. 2003; Leroy et al. 2009).
There is no apparent associated nebulosity (Testor \& Lortet 1987), however, and comparisons of the strengths and velocities of the interstellar emission and absorption features in this sight line indicate that much of the material (both neutral and ionized) seen in emission is located behind the star (Welty et al. 2012 and in preparation). 
The mild depletions of Mg, Si, and Ti seen for several of the SMC components -- compared to the severe depletions of Fe and Ni -- are indicative of depletion patterns unlike those seen in the local Galactic ISM (Welty et al. 2001 and in preparation; Welty \& Crowther 2010; cf. Sofia et al. 2006).}
\item{Sk$-$67~5 is located in a diffuse \ion{H}{2} region (N3, DEM~7) in the northwestern part of the LMC.
Friedman et al. (2000) and Andr\'{e} et al. (2004) have discussed the interstellar absorption seen in early {\it FUSE} and optical spectra; analysis of the higher-resolution STIS spectra indicates that most of the interstellar components exhibit fairly uniform, relatively mild depletions (Welty et al. 2004 and in preparation; Welty \& Crowther 2010).}
\item{Sk$-$68~73 is a member of the LH~49 association, in the southern part of the N44 complex of \ion{H}{2} regions and molecular gas located in the central part of the LMC, and appears to be responsible for ionizing the N44H \ion{H}{2} region (Megnier et al. 1996; Kim et al. 1998; Chen et al. 2009).
The sight line lies within the contours of CO emission in the recent MAGMA survey of the LMC (Wong et al. 2011), and CO absorption is detected in the {\it FUSE} spectrum (Welty et al., in preparation).
The overall depletion of Ti is slightly more severe than for several other LMC sight lines with similar $f$(H$_2$) (Welty \& Crowther 2010).}
\item{Sk$-$70~115, the most luminous star in the LH~117 association (Lucke \& Hodge 1970; Massey et al. 1989), is located near the edge of the LMC2 region (southeast of 30 Dor), and appears to be associated with the N180 \ion{H}{2} region. 
Other stars in that region (including the nearby Sk$-$70~116) exhibit UV extinction curves with a weak 2175~\AA\ bump and relatively steep far-UV rise (Gordon et al. 2003).
The various interstellar components exhibit marked differences in Ti depletion -- severe in the main 220 km~s$^{-1}$ component, but mild in many of the components at higher velocities (Caulet \& Newell 1996; Welty \& Crowther 2010); the relatively mild depletions of Mg and Si in the 220 km~s$^{-1}$ component are reminiscent of the unusual depletion patterns seen toward Sk~155 (Welty et al. 2004 and in preparation).}
\end{itemize}

\subsection{Optical spectra}
\label{sec-opt}

The good correlations and nearly linear relationships between the column densities of the trace neutral species \ion{C}{1}, \ion{Na}{1}, and \ion{K}{1} found in the Galactic ISM (Jenkins \& Shaya 1979; Welty \& Hobbs 2001) suggest that those three species are similarly distributed and respond similarly to local physical conditions in the ISM (e.g., density, radiation field) -- and thus that high-resolution and/or high-S/N ratio optical spectra of \ion{Na}{1} and \ion{K}{1} absorption can be used to help interpret the absorption from \ion{C}{1} observed (generally at lower resolution and/or S/N ratio) in the UV.
For the seven sight lines considered in this paper, high-resolution (FWHM = 1.35--2.0 km~s$^{-1}$) spectra of the \ion{Na}{1} D lines near 5890 and 5896 \AA\ were obtained with the ESO 3.6 m telescope and coud\'{e} echelle spectrograph (CES) during runs in 1998 and 2000, while spectra of somewhat lower resolution (FWHM = 4.5--4.9 km~s$^{-1}$) but much higher S/N ratio were obtained for both the \ion{Na}{1} D lines and the weaker \ion{Na}{1} U lines near 3302 \AA\ with the ESO VLT-UT2 telescope and UVES spectrograph in 2003 (Welty et al. 2006; Welty \& Crowther 2010 and in preparation).
As described in detail by Welty \& Crowther (in preparation), standard procedures within {\sc iraf} were used to obtain wavelength-calibrated 1-D extracted spectra from the raw 2-D spectral images.
Normalization of those spectra, measurement of the equivalent widths of the absorption lines, and initial column density estimates based on the apparent optical depths (AOD) in the observed line profiles were accomplished using a locally developed spectral analysis program ({\sc specp}).

\subsection{UV spectra}
\label{sec-uv}

The seven Magellanic Clouds stars included in this study are the only ones for which high-resolution (FWHM $\sim$ 2.7 km~s$^{-1}$) UV spectra covering a number of the \ion{C}{1} multiplets have been obtained.
In each case, use of the STIS E140H echelle grating yielded a spectral segment of about 200 \AA, centered on the nominal wavelength of the setting (1234, 1271, or 1307 \AA) -- thus including the seven \ion{C}{1} multiplets listed in Table~\ref{tab:c1}.\footnotemark
\footnotetext{The \ion{C}{1} multiplets near 1188, 1192, 1193, 1194 \AA\ were not used in this study, as (1) they were not covered in some of the sight lines, (2) where they were covered, the S/N ratios are generally rather low, and (3) there is often significant blending with Galactic and/or Magellanic Clouds absorption from \ion{Cl}{1} ($\lambda$1188), \ion{S}{3} ($\lambda$1190), and/or \ion{Si}{2} ($\lambda$1190, $\lambda$1193).}
As the targets are all relatively faint (and somewhat reddened), the total exposure times for those E140H settings ranged from about 2.8 hours for Sk$-$67~5 ($V$ $\sim$ 11.3) to about 17.7 hours for Sk~13 ($V$ $\sim$ 12.5).

\begin{figure}
\epsscale{0.8}
\plotone{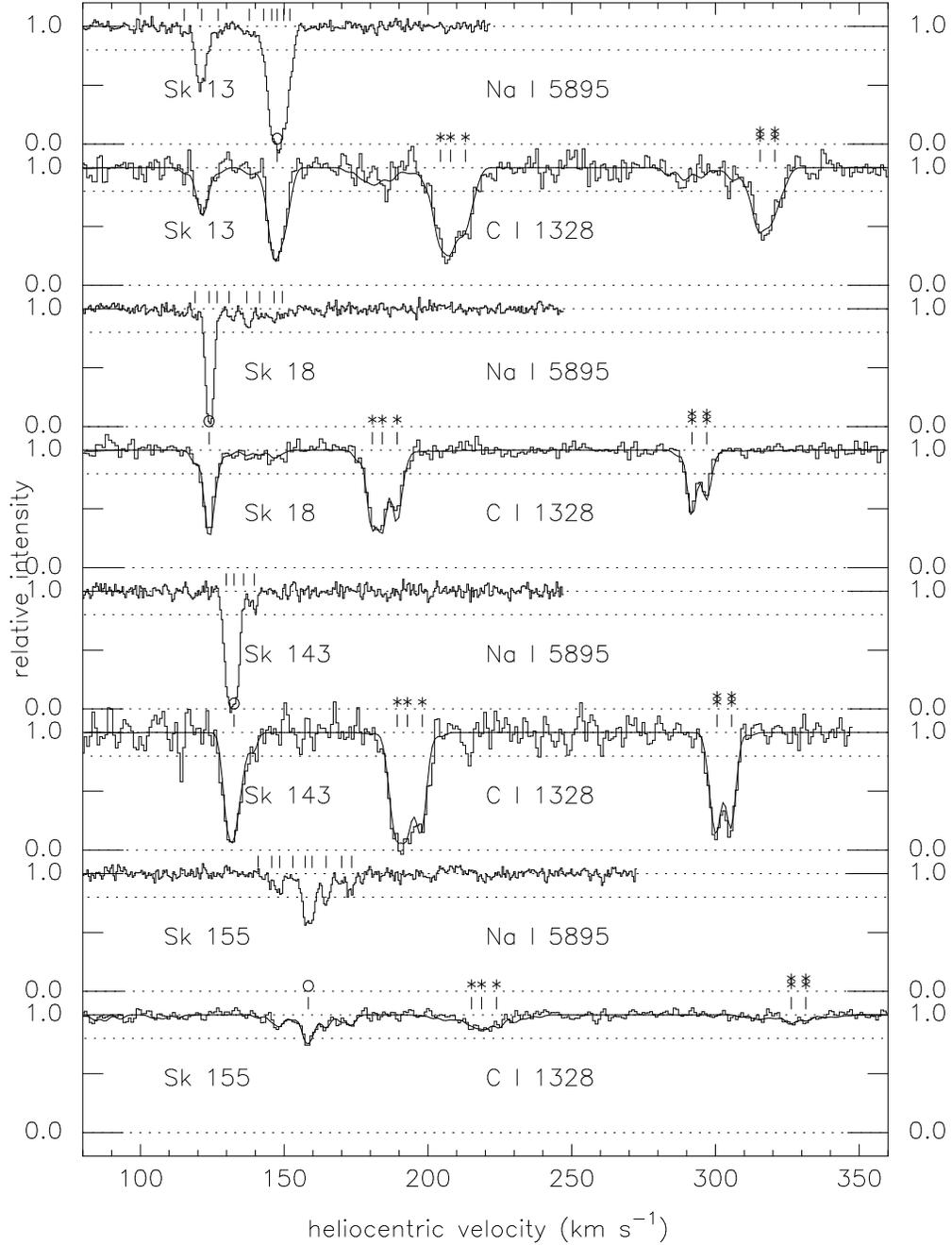}
\caption{High-resolution optical spectra (FWHM = 1.35 km~s$^{-1}$) of \ion{Na}{1} $\lambda$5895 and STIS E140H spectra (FWHM = 2.7 km~s$^{-1}$) of \ion{C}{1} $\lambda$1328 for four SMC stars.
Only the absorption from gas in the SMC, at $v$ $>$ 80 km~s$^{-1}$, is shown.
The histograms give the normalized observed spectra; the smooth curves for \ion{C}{1} give the theoretical profiles for the adopted component structure (Table~\ref{tab:comps}).
The tick marks above the \ion{Na}{1} profiles denote the individual components found in the detailed fits.
The tick marks above the \ion{C}{1} profiles give the locations of the ground and excited fine-structure state lines corresponding to the strongest \ion{Na}{1} component in each sight line.
The absorption from the excited \ion{C}{1} states (* or $\stackrel{\textstyle *}{*}$) is seen (in blends) at about 60 and 170 km~s$^{-1}$ beyond the ground state absorption (o), which is aligned with \ion{Na}{1}.}
\label{fig:nacs}
\end{figure}

\begin{figure}
\epsscale{0.8}
\plotone{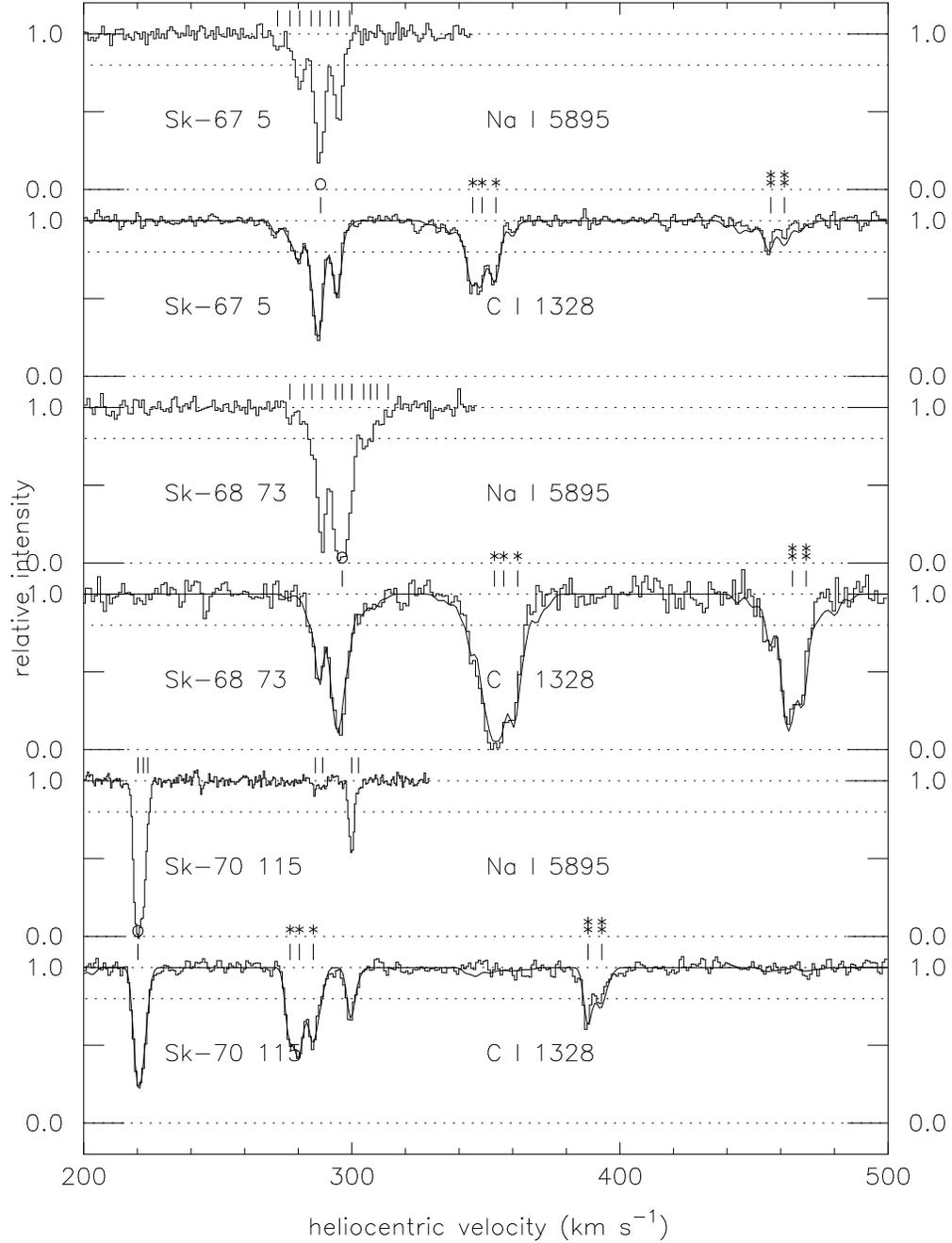}
\caption{High-resolution optical spectra (FWHM = 1.35--2.0 km~s$^{-1}$) of \ion{Na}{1} $\lambda$5895 and STIS E140H spectra (FWHM = 2.7 km~s$^{-1}$) of \ion{C}{1} $\lambda$1328 for three LMC stars.
Only the absorption from gas in the LMC, at $v$ $>$ 200 km~s$^{-1}$, is shown.
The histograms give the normalized observed spectra; the smooth curves for \ion{C}{1} give the theoretical profiles for the adopted component structure (Table~\ref{tab:comps}).
The tick marks above the \ion{Na}{1} profiles denote the individual components found in the detailed fits.
The tick marks above the \ion{C}{1} profiles give the locations of the ground and excited fine-structure state lines corresponding to the strongest \ion{Na}{1} component in each sight line.
The absorption from the excited \ion{C}{1} states (* or $\stackrel{\textstyle *}{*}$) is seen (in blends) at about 60 and 170 km~s$^{-1}$ beyond the ground state absorption (o; which is aligned with \ion{Na}{1}).}
\label{fig:nacl}
\end{figure}

The default pipeline-extracted spectra were obtained from the MAST archive{\footnotemark}, and the multiple individual spectra for each sight line were combined via sinc interpolation to a common, slightly over-sampled heliocentric wavelength grid.
\footnotetext{http://archive.stsci.edu/hst}
Because of the relatively low S/N ratios characterizing the individual E140H spectra, it was generally difficult to discern any systematic relative velocity offsets among the individual spectra -- so no offsets were applied in combining the spectra.
(Where they can be measured, however, such offsets are typically $\la$ 0.5 km~s$^{-1}$ for STIS E140H spectra.)
Spectral segments encompassing each of the \ion{C}{1} multiplets were normalized via polynomial fits to the line-free regions in each segment; equivalent widths and AOD column density estimates were measured for the unblended lines in those normalized spectra.
The S/N ratios (per resolution element) characterizing the spectra, gauged from fluctuations in the continuum regions near the \ion{C}{1} multiplets, range from about 14--17 for Sk~143 to about 46--60 for Sk~155.
The normalized spectra of the \ion{C}{1} multiplet near 1328~\AA\ are shown, together with the corresponding spectra of \ion{Na}{1} $\lambda$5895, in Figures~\ref{fig:nacs} (SMC) and \ref{fig:nacl} (LMC); the profiles of the ground state \ion{C}{1} line and the \ion{Na}{1} line are strikingly similar.
The spectra of all of the \ion{C}{1} multiplets analyzed here (both Galactic and Magellanic Clouds absorption) are shown in Appendix Figures~\ref{fig:sk13c1} through \ref{fig:sk70d115c1}.

\begin{deluxetable}{crrrrrr}
\tablecolumns{7}
\tabletypesize{\scriptsize}
\tablecaption{C I Multiplets \label{tab:c1}}
\tablewidth{0pt}

\tablehead{
\multicolumn{1}{c}{Multiplet}&
\multicolumn{2}{c}{C I}&
\multicolumn{2}{c}{C I*}&
\multicolumn{2}{c}{C I**}\\
\multicolumn{1}{c}{ }&
\multicolumn{1}{c}{$\lambda$}&
\multicolumn{1}{c}{$f$}&
\multicolumn{1}{c}{$\lambda$}&
\multicolumn{1}{c}{$f$}&
\multicolumn{1}{c}{$\lambda$}&
\multicolumn{1}{c}{$f$}}

\startdata
9    & 1260.7351 & 0.05880 & 1260.9262 & 0.02608 & 1261.4255 & 0.02071 \\ 
     &           &         & 1260.9961 & 0.02204 & 1261.5519 & 0.04417 \\
     &           &         & 1261.1224 & 0.02731 &           &         \\
8.01 & 1270.1430 & 0.00216 & 1270.4081 & 0.00041 & 1270.8440 & 1.45e-8 \\ 
7.01 & 1276.4822 & 0.01180 & 1276.7498 & 0.00824 & 1277.1900 & 0.00075 \\ 
7    & 1277.2452 & 0.13144 & 1277.2827 & 0.08142 & 1277.5501 & 0.08622 \\ 
     &           &         & 1277.5131 & 0.03950 & 1277.7233 & 0.02764 \\
     &           &         &           &         & 1277.9539 & 0.00307 \\
6    & \nodata   & \nodata & 1279.0562 & 0.00610 & 1279.2290 & 0.00900 \\
     &           &         &           &         & 1279.4980 & 0.00362 \\
5    & 1280.1352 & 0.04806 & 1279.8907 & 0.02817 & 1280.3331 & 0.02759 \\ 
     &           &         & 1280.4043 & 0.01320 & 1280.8471 & 0.01329 \\
     &           &         & 1280.5975 & 0.01777 &           &         \\
4    & 1328.8333 & 0.08985 & 1329.0849 & 0.03601 & 1329.5775 & 0.05785 \\ 
     &           &         & 1329.1004 & 0.04350 & 1329.6004 & 0.03064 \\
     &           &         & 1329.1233 & 0.03086 &           &         \\
\enddata
\tablecomments{All $f$-values are from Jenkins \& Tripp (2001, 2011) except for $\lambda$1270.8440, which is from Morton (2003).}
\end{deluxetable}

\clearpage

\clearpage

\subsection{Fits to line profiles}
\label{sec-fits}

As seen for the local Galactic ISM, sight lines probing the ISM of the SMC and LMC can exhibit complex structure, even when observed at only moderately high spectral resolution (FWHM $\sim$ 5 km~s$^{-1}$; e.g., Wayte 1990; Vladilo et al. 1993; van Loon et al. 2013); higher resolution spectra (FWHM $\la$ 3 km~s$^{-1}$) typically reveal numerous narrow, closely-spaced components (Andreani et al. 1987; Pettini \& Gillingham 1988; Welty et al. 1999a, 2006; Welty \& Crowther, in preparation).
Multicomponent fits to the line profiles of the various \ion{C}{1} multiplets were therefore used to obtain the column densities of the ground and excited fine-structure levels of \ion{C}{1} for the components discernible in our seven sight lines, using the program {\sc fits6p} (e.g., Welty et al. 2003) and a variant of that program which can fit multiple, widely separated lines having a common component structure.
Such fits explicitly account for the blending between lines from different levels and between different interstellar components, as well as for possible saturation effects in the stronger absorption features.
Potential disadvantages of this approach are the assumption of symmetric Voigt profiles for each component and the somewhat subjective nature of deciding how many components are present.\footnotemark
\footnotetext{While the F-test has sometimes been used to determine how many components are required in such fits, the F-test appears not to be appropriate for that particular decision (Protassov et al. 2002).}

In view of the generally good correlation between the column densities of \ion{C}{1}$_{\rm tot}$\footnotemark\ and \ion{Na}{1} and the apparent similarity of their line profiles (Figures~\ref{fig:nacs} and \ref{fig:nacl}), fits to the highest resolution \ion{Na}{1} D-line profiles (Welty \& Crowther, in preparation) were used to determine both the number of components and the widths ($b$-values) and velocities of those components for the trace neutral species in each of the sight lines.
\footnotetext{While the term \ion{C}{1} in the text generally refers to neutral carbon (in whatever state), we follow JT11 in designating the specific, individual column densities of the ground and excited fine-structure states as $N$(\ion{C}{1}), $N$(\ion{C}{1}*), and $N$(\ion{C}{1}**), and of the total column density of \ion{C}{1} (in all states) as $N$(\ion{C}{1}$_{\rm tot}$).}
In order to facilitate convergence, most of the component $b$-values were held fixed (at values determined by manual adjustment) in the final iterative fits to the profiles.
The $b$-values and column densities of the strongest \ion{Na}{1} components were further constrained by requiring consistent fits for both the strong D lines (in both CES and UVES spectra) and the much weaker U lines near 3302 \AA.
The adopted \ion{Na}{1} component structures for these SMC and LMC sight lines are given in Appendix Table~\ref{tab:comps}.
Toward the SMC, components at (heliocentric) velocities less than about 80 km~s$^{-1}$ are assumed to arise in the Galactic disk and halo, while components at higher velocities arise in the SMC.
Toward the LMC, the dividing point between Galactic and LMC components is assumed to be at about 100 km~s$^{-1}$ -- though it is rare to detect absorption from \ion{Na}{1} in the intermediate-velocity gas between 100 and 200 km~s$^{-1}$ (e.g., Lehner et al. 2009; Smoker et al. 2015; Welty \& Crowther, in preparation).
The SMC and LMC component \ion{Na}{1} $b$-values generally lie between 0.4 and 1.2 km~s$^{-1}$ -- similar to the range found for most Galactic \ion{Na}{1} components from fits to higher-resolution spectra (Welty et al. 1994; Welty \& Crowther, in preparation).
The slightly higher median $b$-values for this small sample -- about 0.8 km~s$^{-1}$ for the Galactic components and 1.0 km~s$^{-1}$ for the Magellanic Clouds components, versus the 0.7 km~s$^{-1}$ found by Welty et al. (1994) -- can probably be attributed to the lower resolution and modest S/N ratios characterizing the present spectra.

The component structure derived for \ion{Na}{1} was then used to fit the various \ion{C}{1} lines from both ground and excited fine-structure levels -- varying only the column densities of the individual components for each level and the slight overall velocity offsets (generally less than 0.5 km~s$^{-1}$) between the profiles of the different multiplets.
An instrumental width slightly larger than the nominal 2.7 km~s$^{-1}$ was adopted, to account for the effects of slight possible (uncorrected) offsets among the individual spectra contributing to the overall sums.
Simultaneous fits were performed both to the unblended lines from several multiplets (separately for each level) and to all the lines (from all three levels) for each individual multiplet.
As for \ion{Na}{1}, it was particularly valuable to have spectra for \ion{C}{1} lines characterized by a wide range in intrinsic strength (spanning a factor of $\sim$60 in $f\lambda$ for the ground state lines of the various multiplets; Table~\ref{tab:c1}).
Given that range in line strength, it was possible to derive reasonably well constrained values for both the $b$-values and the column densities of the strongest \ion{C}{1} components, by requiring consistent column densities for all the multiplets.
The resulting adopted $b$-values for the strongest \ion{C}{1} components are typically $\sim$ 0.1--0.2 km~s$^{-1}$ larger than those found for the corresponding components in \ion{Na}{1} -- consistent with the smaller atomic weight of carbon, but perhaps also suggestive of a slightly broader distribution for \ion{C}{1} than for \ion{Na}{1}.
The column densities of the weaker \ion{C}{1} components are relatively insensitive to the choice of $b$.
The derived component column densities for the ground and excited fine-structure states of \ion{C}{1} are listed in Appendix Table~\ref{tab:comps}; the line profiles for those adopted column densities are shown as the smooth curves in Appendix Figures~\ref{fig:sk13c1} through \ref{fig:sk70d115c1}.
The profiles computed for the weaker $\lambda$1270 and $\lambda$1276 multiplets indicate that the ground state column densities for the main components in each sight line cannot be much larger than the adopted values listed in Table~\ref{tab:comps} -- so that the excitation and thermal pressures estimated for those components (below) are unlikely to be overestimated.
The total Magellanic Clouds column densities for H, H$_2$, \ion{C}{1}$_{\rm tot}$, \ion{Na}{1}, \ion{K}{1}, and CH are listed in Table~\ref{tab:mc}.

\begin{deluxetable}{lrrcrrrrr}
\tablecolumns{9}
\tabletypesize{\scriptsize}
\tablecaption{Magellanic Clouds Column Densities: H, H$_2$, \ion{C}{1}$_{\rm tot}$, \ion{Na}{1}, \ion{K}{1}, CH \label{tab:mc}}
\tablewidth{0pt}

\tablehead{
\multicolumn{1}{c}{Star}&
\multicolumn{1}{c}{$N$(H)}&
\multicolumn{1}{c}{$N$(H$_2$)}&
\multicolumn{1}{c}{log[$f$(H$_2$)]}&
\multicolumn{1}{c}{$T_{01}$}&
\multicolumn{1}{c}{$N$(C~I$_{\rm tot}$)}&
\multicolumn{1}{c}{$N$(Na~I)}&
\multicolumn{1}{c}{$N$(K~I)}&
\multicolumn{1}{c}{$N$(CH)}}

\startdata
Sk 13      & 22.04$\pm$0.02 & 20.36$\pm$0.07 & $-$1.40 & 66 & 14.06$\pm$0.03 & 12.81$\pm$0.05 & 11.38$\pm$0.02 & $<$12.33 \\
Sk 18      & 21.70$\pm$0.05 & 20.63$\pm$0.05 & $-$0.84 & 53 & 14.09$\pm$0.02 & 12.98$\pm$0.10 & 11.20$\pm$0.10 & 12.08$\pm$0.15 \\
Sk 143     & 21.00$\pm$0.05 & 20.93$\pm$0.09 & $-$0.20 & 45 & 14.93$\pm$0.06 & 13.72$\pm$0.06 & 12.65$\pm$0.10 & 13.54$\pm$0.04 \\
Sk 155     & 21.42$\pm$0.05 & 19.15$\pm$0.10 & $-$1.97 & 82 & 13.29$\pm$0.03 & 11.94$\pm$0.04 & \nodata        & $<$12.26 \\
\hline
Sk$-$67~5  & 21.00$\pm$0.05 & 19.46$\pm$0.05 & $-$1.26 & 57 & 13.89$\pm$0.02 & 12.58$\pm$0.15 & 11.04$\pm$0.04 & $<$12.28 \\
Sk$-$68~73 & 21.60$\pm$0.04 & 20.09$\pm$0.20 & $-$1.24 & 57 & 14.43$\pm$0.02 & 13.38$\pm$0.10 & 11.76$\pm$0.04 & 12.80$\pm$0.05 \\
Sk$-$70~115& 21.30$\pm$0.05 & 19.94$\pm$0.07 & $-$1.10 & 53 & 13.89$\pm$0.03 & 12.71$\pm$0.05 & \nodata        & 11.96$\pm$0.15 \\
\enddata
\tablecomments{Column densities (cm$^{-2}$) are logarithmic.
Values for H and H$_2$ are from Welty et al. 2012 (where original references are given); values for C~I$_{\rm tot}$ are from this paper; values for Na~I and K~I are from Welty \& Crowther, in preparation; values for CH are from Welty et al. 2006.
Uncertainties are approximately 1$\sigma$; limits are 3$\sigma$.}
\end{deluxetable}

\clearpage

\subsection{\ion{C}{1} $f$-values}
\label{sec-c1f}

JT11 found systematic differences between the $f$-values derived in their global analyses of the \ion{C}{1} multiplets seen in STIS spectra and the experimental or theoretical $f$-values reported in the literature (e.g., Wiese et al. 1996; Morton 2003; Froese Fischer 2006) -- with progressively larger differences for the weaker multiplets.
JT11 explored various possible explanations for the differences -- primarily related to possible saturation effects for narrow, unresolved components -- but found no resolution.
We have compared empirical curves of growth, derived from fits to high-resolution spectra of \ion{Na}{1} and \ion{K}{1} (Welty et al. 1994 and in preparation; Welty \& Hobbs 2001), with the equivalent widths of unblended \ion{C}{1}, \ion{C}{1}*, and \ion{C}{1}** lines seen toward X~Per, 23~Ori, HD~62542, and $\rho$~Oph~A (Zsarg\'{o} et al. 1997; Welty et al. 1999b and in preparation), for the $f$-values listed in Morton (2003), Froese Fischer (2006), and JT11.
While the stronger lines (with the best-determined $f$-values) generally lie on the flat part of the empirical curves of growth -- so that there is some uncertainty in matching the points to the curves -- the comparisons generally suggest that the $f$-values for the weaker lines are too low in Morton (2003) and Froese Fischer (2006), but may be too high in JT11.
The ''true'' $f$-values for those weaker lines thus may lie between the low and high values given in those references -- consistent with hints from the slopes of the relationships between the column densities of \ion{C}{1}$_{\rm tot}$, \ion{Na}{1}, and \ion{K}{1} (discussed below).
Accurate experimental determinations of the $f$-values for those weak \ion{C}{1} lines would be quite valuable.

While such $f$-value uncertainties obviously affect the determination of \ion{C}{1} column densities, they can also affect analyses of the excitation.
For example, at relatively low thermal pressures (as are common in the local Galactic ISM), the ground state populations are much higher than those in the excited states, and the ground state lines in a given multiplet will be more subject to saturation effects than the corresponding excited state lines in that multiplet.
In some cases, the ground and excited state populations may best be determined from different multiplets -- but only if the relative $f$-values are accurately known.
Fortunately, the relative populations of the \ion{C}{1} excited states appear to be higher in the SMC and LMC (as discussed below) -- reducing any differential saturation effects within each individual multiplet.
The \ion{C}{1} excitation derived for these SMC and LMC sight lines thus will not be as sensitive to the choice of $f$-values as it might be for local Galactic sight lines.
We have adopted the JT11 $f$-values for this study of the SMC and LMC -- as that choice facilitates comparisons with the \ion{C}{1} abundances and excitation that JT11 obtained for Galactic sight lines.

\subsection{$^{12}$C/$^{13}$C ratio}
\label{sec-1213}

In principle, contributions from the less abundant isotope $^{13}$C might affect the profiles of the strongest \ion{C}{1} velocity components.
In the local Galactic ISM, the $^{12}$C/$^{13}$C ratio is $\sim$ 70 (e.g., Sheffer et al. 2007; Ritchey et al. 2011; and references therein); the ratio is expected to be somewhat higher for less evolved systems (e.g., Prantzos et al. 1996).
Determination of that ratio from different molecular species (e.g., $^{12}$CO/$^{13}$CO, $^{12}$CN/$^{13}$CN) can be complicated by fractionation effects, however -- depending on factors such as the energetics of the chemical reactions involved in the creation and destruction of the isotopic variants and possible self-shielding against photodissociation (for the more abundant variants).
Because CH$^+$ must be formed either non-thermally or at relatively high temperatures, however, the $^{12}$CH$^+$/$^{13}$CH$^+$ ratio should not be significantly affected by fractionation -- and so should provide good estimates for the $^{12}$C/$^{13}$C ratio (e.g., Ritchey et al. 2011).
Examination of the available UVES spectra for several of the LMC sight lines with detections of the strongest $\lambda$4232 line of CH$^+$ (Welty et al. 2006) yields only upper limits for $N$($^{13}$CH$^+$), which imply corresponding 3$\sigma$ lower limits on the $^{12}$C/$^{13}$C ratio of about 4 (Sk$-$68~73), 10 (LH10-3061), and 20 (Sk$-$67~2).
Observations of emission from a number of other molecular species at mm wavelengths, in the N113 star-forming region and the N159 complex of \ion{H}{2} regions in the LMC (Johansson et al. 1994; Wang et al. 2009) and in the N27 compact \ion{H}{2} region in the SMC (Heikkil\"{a} et al. 1999), have suggested $^{12}$C/$^{13}$C ratios of order 50.

For \ion{C}{1}, the lines of $^{13}$\ion{C}{1} are shifted by only about 0.65, $-$3.30, and $-$1.65 km~s$^{-1}$, relative to those of $^{12}$\ion{C}{1}, for the multiplets near 1656, 1560, and 1328 \AA, respectively (Morton 2003); similar data were not provided for other \ion{C}{1} multiplets.
While Carswell et al. (2011) note a possible detection of $^{13}$\ion{C}{1} in a component of the $z$ = 1.776 damped Ly$\alpha$ system toward the QSO 1331+170, with $^{12}$C/$^{13}$C $\sim$ 20, that weak contribution to the \ion{C}{1} absorption had little effect on the inferred excitation of \ion{C}{1}.
If $^{12}$C/$^{13}$C $\ga$ 50 in the Magellanic Clouds, then $^{13}$\ion{C}{1} would only be (barely) detectable for the several strongest components in the seven sight lines considered in this study, and it is unlikely that the derived relative level populations would be significantly affected.

\begin{figure}
\epsscale{0.9}
\plotone{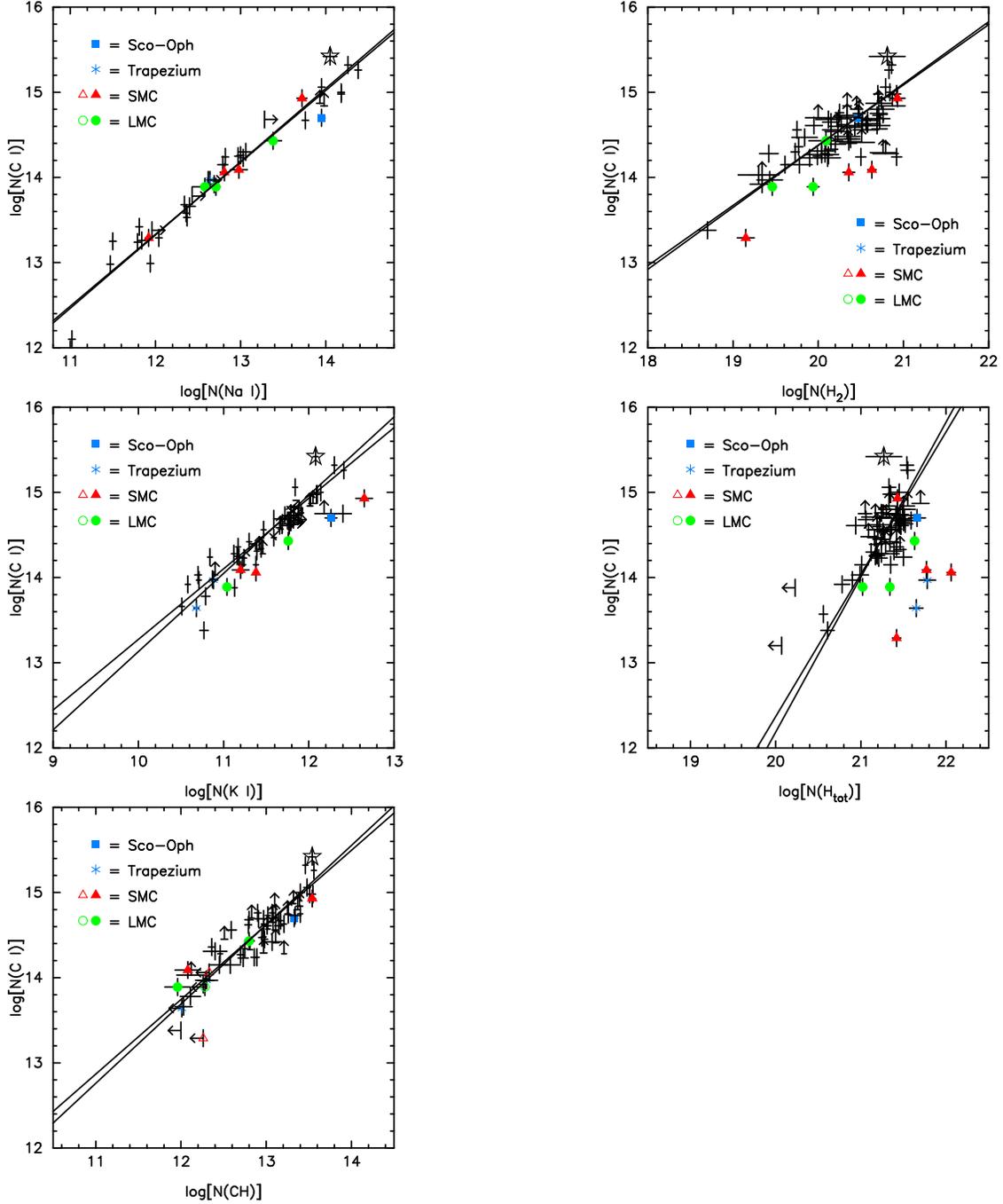}
\caption{Column density of \ion{C}{1}$_{\rm tot}$ vs. column densities of \ion{Na}{1}, \ion{K}{1}, CH, H$_2$, and H$_{\rm tot}$ (see Appendix Table~\ref{tab:gal}).
Red triangles denote SMC sight lines and green circles denote LMC sight lines (\ion{C}{1} values from this study).
All others are Galactic, with most $N$(\ion{C}{1}$_{\rm tot}$) values from Burgh et al. (2010) or JT11; seven of the lowest $N$(\ion{C}{1}$_{\rm tot}$) vs. $N$(\ion{Na}{1}) are the Galactic values toward our SMC and LMC targets.
The large open star is for HD~62542, which has the largest $N$(\ion{C}{1}) in our Galactic sample.
The solid lines represent the best weighted and unweighted fits to the Galactic points (not including the Sco-Oph and Trapezium sight lines).
The trends with \ion{Na}{1}, \ion{K}{1}, and CH are quite tight, with $r$ = 0.91--0.98; the slopes are $\sim$ 0.85--0.90 in those three cases.}
\label{fig:c1abund}
\end{figure}

\begin{deluxetable}{lcrccc}
\tablecolumns{6}
\tabletypesize{\scriptsize}
\tablecaption{\ion{C}{1} versus Other Species (Galactic ISM) \label{tab:corr}}
\tablewidth{0pt}

\tablehead{
\multicolumn{1}{c}{X}&
\multicolumn{1}{c}{$r$\tablenotemark{a}}&
\multicolumn{1}{c}{A}&
\multicolumn{1}{c}{B}&
\multicolumn{1}{c}{rms\tablenotemark{b}}&
\multicolumn{1}{c}{N\tablenotemark{c}}}

\startdata
Na I         & 0.978 & 3.20$\pm$0.52 & 0.84$\pm$0.04 & 0.14 & 21 \\
             &       & 3.00$\pm$0.52 & 0.86$\pm$0.04 & 0.14 & 21 \\
K I          & 0.916 & 4.98$\pm$0.63 & 0.83$\pm$0.06 & 0.12 & 33 \\
             &       & 3.95$\pm$0.71 & 0.92$\pm$0.06 & 0.14 & 33 \\
CH           & 0.906 & 3.23$\pm$0.84 & 0.88$\pm$0.06 & 0.12 & 43 \\
             &       & 2.52$\pm$0.82 & 0.93$\pm$0.06 & 0.12 & 43 \\
H$_2$        & 0.812 & 0.23$\pm$1.18 & 0.71$\pm$0.06 & 0.16 & 50 \\
             &       &$-0.16\pm$1.17 & 0.73$\pm$0.06 & 0.16 & 50 \\
H$_{\rm tot}$& 0.721 &$-21.06\pm3.39$& 1.67$\pm$0.16 & 0.14 & 47 \\
             &       &$-24.20\pm3.90$& 1.82$\pm$0.18 & 0.16 & 47 \\
\enddata
\tablecomments{The two lines for each species are for weighted and unweighted fits: log[$N$(C~I$_{\rm tot}$)] = A + B $\times$ log[$N$(X)].}
\tablenotetext{a}{The linear correlation coefficient}
\tablenotetext{b}{Root mean square distance of points from best-fit line}
\tablenotetext{c}{Number of sight lines in the sample}
\end{deluxetable}

\section{RESULTS}
\label{sec-res}

\subsection{Abundance and behavior of \ion{C}{1}}
\label{sec-c1ab}

In general, the abundances of trace neutral species \ion{X}{1} should depend similarly on both the overall strength of the UV radiation field (responsible for photoionization) and the local temperature and density (which affect the corresponding radiative, dielectronic, and grain-assisted recombination processes).
Some differences may arise for individual clouds, however, due to differences in the wavelength dependence of the photoionization cross sections (if the shape of the UV field is very different) and/or in the depletion behavior of each element.
Previous comparisons of the abundances of \ion{Li}{1}, \ion{C}{1}$_{\rm tot}$, \ion{Na}{1}, and \ion{K}{1} in the Galactic ISM, for example, have found roughly linear relationships between the column densities of those species (e.g., Jenkins \& Shaya 1979; Welty \& Hobbs 2001); the shallower slopes for \ion{Fe}{1} and \ion{Ca}{1} (versus \ion{K}{1}) can plausibly be attributed to the increasingly severe depletion of iron and calcium in the higher column density sight lines (Welty et al. 2003).
The molecular species CH and H$_2$ [for $N$(H$_2$) $\ga$ 10$^{18}$ cm$^{-2}$] should have similar dependences on the UV field (for photodissociation) and local density (for formation); their column densities also exhibit roughly linear correlations with $N$(\ion{Na}{1}) and $N$(\ion{K}{1}) in the Galactic ISM (Welty \& Hobbs 2001; Welty et al. 2006).

Figure~\ref{fig:c1abund} revisits the relationships between $N$(\ion{C}{1}$_{\rm tot}$) and the corresponding column densities of \ion{Na}{1}, \ion{K}{1}, CH, H$_2$, and H$_{\rm tot}$, for the Galactic sight lines included in the recent surveys of Burgh et al. (2010) and JT11 (Appendix Table~\ref{tab:gal}) and for the seven SMC and LMC sight lines examined in this study (Table~\ref{tab:mc}).
The lower end of the relationship between $N$(\ion{C}{1}$_{\rm tot}$) and $N$(\ion{Na}{1}) is largely defined by the Galactic components seen toward our seven SMC and LMC stars -- which are quite consistent with an extrapolation of the trend seen at higher column densities in the \ion{C}{1} data from Burgh et al. (2010) and JT11 and with the values found for several lower column density Galactic sight lines (Jenkins 2002; Meyer et al. 2012).
As noted above, $N$(\ion{C}{1}$_{\rm tot}$) is fairly tightly correlated ($r$ = 0.91--0.98) with $N$(\ion{Na}{1}), $N$(\ion{K}{1}), and $N$(CH) in the local Galactic ISM (Table~\ref{tab:corr}).
Both considerations of ionization equilibrium and relative depletions (as noted above) and the striking similarities in the observed absorption-line profiles of the three trace neutral species (in many sight lines) suggest that those species track each other fairly well in the diffuse ISM -- so that the tight correlations among their column densities are not due just to the averaging of many disparate components over long path lengths.
The slopes of the relationships -- $\sim$0.85--0.90 versus $N$(\ion{Na}{1}), $N$(\ion{K}{1}), and $N$(CH), and $\sim$0.7 versus $N$(H$_2$) -- are somewhat smaller than those obtained previously [e.g., $\sim$1.1--1.3 versus $N$(\ion{Na}{1}) and $N$(\ion{K}{1}); Welty \& Hobbs 2001], however; see also the comparisons shown in Liszt (2011).
The shallower slopes found here likely reflect the use of the JT11 \ion{C}{1} $f$-values -- as the higher $N$(\ion{C}{1}$_{\rm tot}$) values generally are more dependent on observations of the weaker \ion{C}{1} lines, for which the JT11 $f$-values are progressively larger than those used in previous studies.
Unfortunately, there is very little overlap in sight lines between the recent surveys and the earlier studies based on data from {\it Copernicus} and {\it HST}/GHRS (e.g., as collected in Wolfire et al. 2008), and direct comparisons will depend on which \ion{C}{1} lines were used in each case.

In the Galactic ISM, both $N$(\ion{Na}{1}) and $N$(\ion{K}{1}) exhibit nearly quadratic relationships with $N$(H$_{\rm tot}$), consistent with considerations of ionization equilibrium (e.g., Welty \& Hobbs 2001).
In Figure~\ref{fig:c1abund}, the relationship between the Galactic $N$(\ion{C}{1}$_{\rm tot}$) and $N$(H$_{\rm tot}$) is similar to those seen for $N$(\ion{Na}{1}) and $N$(\ion{K}{1}), but (again) with slightly shallower slope (see also Burgh et al. 2010; Liszt 2011).
The lower $N$(\ion{C}{1}$_{\rm tot}$), relative to $N$(H$_{\rm tot}$), seen toward HD~147888 (in Sco-Oph) and toward the Orion Trapezium region stars HD~37021 and HD~37061 is likely due to increased photoionization of neutral carbon in the stronger-than-average radiation fields in those regions (consistent with JT11's analysis of the \ion{C}{1} excitation) -- as seen also for $N$(\ion{Na}{1}) and $N$(\ion{K}{1}) there (Welty \& Hobbs 2001).
The strong \ion{C}{1} seen toward HD~62542, which at log $N$(\ion{C}{1}$_{\rm tot}$) $\sim$ 15.4 lies slightly above the mean trends in all five correlation plots, likely reflects the very steep far-UV extinction in that sight line, which preferentially inhibits the photoionization of neutral carbon relative to that of other trace neutral species with lower ionization potentials (Cardelli \& Savage 1988;  Welty et al., in preparation).

In the LMC and SMC, the characteristic sub-Solar metallicities, correspondingly lower dust-to-gas ratios (compared to the local Galactic ISM), and typically somewhat enhanced UV radiation fields all should contribute to lower abundances for the various trace neutral species, relative to $N$(H$_{\rm tot}$).
Those overall environmental factors are likely to have similar effects on the abundances of each trace species, however, so that their ratios [$N$(\ion{X}{1})/$N$(\ion{Y}{1})] might not differ much from the typical Galactic values.
[The abundances of CH, H$_2$, and the (unknown) carriers of the diffuse interstellar bands appear to have a more complex behavior in the Magellanic Clouds -- with dependences on those global factors, on consequent differences in cloud structure, and also on local physical conditions (Tumlinson et al. 2002; Cox et al. 2006, 2007; Cox \& Spaans 2006; Welty et al. 2006, 2012, 2013; van Loon et al. 2013; Bailey et al. 2015; see also Vos et al. 2011).]

The plots in Figure~\ref{fig:c1abund} indicate that our seven Magellanic Clouds sight lines do agree well with the Galactic trends for $N$(\ion{C}{1}$_{\rm tot}$) versus $N$(\ion{Na}{1}) and $N$(CH), but that they fall slightly below the Galactic trends for $N$(\ion{C}{1}$_{\rm tot}$) versus $N$(\ion{K}{1}) and $N$(H$_2$).\footnotemark
\footnotetext{$N$(\ion{K}{1}) appears to be slightly less depressed (relative to the column densities of \ion{C}{1}, \ion{Na}{1}, H$_2$, and H$_{\rm tot}$) in both the Sco-Oph region and the Magellanic Clouds (Welty \& Hobbs 2001; Welty \& Crowther, in preparation) -- but not in the Trapezium region (so the difference for \ion{K}{1} appears not to be an effect of the strength of the ambient radiation field).}
And, as has been noted for $N$(\ion{Na}{1}), $N$(\ion{K}{1}), and the equivalent widths of some of the diffuse interstellar bands (Welty \& Hobbs 2001; Cox et al. 2006, 2007; Welty et al. 2006), $N$(\ion{C}{1}$_{\rm tot}$) is generally much lower, relative to $N$(H$_{\rm tot}$), in the LMC and (especially) the SMC than in the local Galactic ISM, by factors larger than the differences in metallicity.
[For comparison, $N$(\ion{Na}{1}) (with a larger sample of measurements) is typically lower by factors of order 30 and 100 in the LMC and SMC (respectively), compared to the values obtained for Galactic sight lines with similar $N$(H$_{\rm tot}$) (Welty \& Crowther, in preparation).]
The one (known) exception is the unusual SMC sight line toward Sk~143, where \ion{C}{1} -- like the other trace neutral and molecular species -- is more consistent with the Galactic relationships (Howk et al. 2012; Welty et al. 2013).
The slight general weakness of $N$(\ion{C}{1}$_{\rm tot}$), relative to $N$(H$_2$), in the LMC and SMC may largely reflect the differences in metallicity.
The more severe deficiencies of all the trace neutral species, relative to $N$(H$_{\rm tot}$), in the Magellanic Clouds appear to be due to the combined effects of the lower metallicities, generally stronger radiation fields, and less grain-assisted recombination (given the lower dust-to-gas ratios and likely smaller fraction of negatively charged small grains) (Welty \& Hobbs 2001; Welty \& Crowther, in preparation; see also Weingartner \& Draine 2001; Liszt 2003).
In addition, carbon appears to be less abundant than most other heavy elements in the LMC and SMC, relative to Solar abundances, by factors of order 1.4--1.8 (Appendix Table~\ref{tab:crefab}).
For \ion{C}{1}, with $\chi_{\rm ion}$ = 11.26 eV, however, the effect of that slightly lower carbon abundance may be offset somewhat by the steeper far-UV extinction in some LMC and most SMC sight lines (e.g., Gordon et al. 2003).

\clearpage

\begin{deluxetable}{lrrccccccrrccr}
\tablecolumns{14}
\tabletypesize{\scriptsize}
\tablecaption{\ion{C}{1} Excitation and Thermal Pressures \label{tab:c1ex}}
\tablewidth{0pt}

\tablehead{
\multicolumn{1}{c}{Star}&
\multicolumn{1}{c}{$v$\tablenotemark{a}}&
\multicolumn{1}{c}{$N$(C~I$_{\rm tot}$)}&
\multicolumn{1}{c}{$N$(C~II)\tablenotemark{b}}&
\multicolumn{1}{c}{$f_1$}&
\multicolumn{1}{c}{$f_2$}&
\multicolumn{5}{c}{- - - - - - - - - C I\tablenotemark{c} - - - - - - - - -}&
\multicolumn{3}{c}{- - - - - H$_2$\tablenotemark{d} - - - - -}\\
\multicolumn{6}{c}{ }&
\multicolumn{2}{c}{log($p/k$)$_{\rm low}$}&
\multicolumn{1}{c}{$g_{\rm low}$}&
\multicolumn{1}{c}{$n_{\rm H}$}&
\multicolumn{1}{c}{I$_{\rm UV}$}&
\multicolumn{1}{c}{$T_{01}$}&
\multicolumn{1}{c}{log($p/k$)}&
\multicolumn{1}{c}{I$_{\rm UV}$}\\
\multicolumn{6}{c}{ }&
\multicolumn{1}{c}{MW}&
\multicolumn{1}{c}{MC}&
\multicolumn{4}{c}{ }&
\multicolumn{1}{c}{MC}&
\multicolumn{1}{c}{MC}}

\startdata
Sk 13       & 121.3 & 13.12$\pm$0.04 &[16.12]& 0.23$\pm$0.05 & 0.14$\pm$0.06 &      & 3.56 & 0.75 &  55 & 1.0 & 66 \\
            & 145.7 & 13.51$\pm$0.09 &[16.51]& 0.30$\pm$0.13 & 0.24$\pm$0.08 &      & 3.74 & 0.55 &  83 & 1.3 & 66 \\
            & 147.6 & 13.54$\pm$0.05 &[16.54]& 0.49$\pm$0.06 & 0.24$\pm$0.06 &      & 4.82 &[1.00]&1000 &[5.0]& 66 & 4.59 & 28 \\ 
            & 149.9 & 13.31$\pm$0.05 &[16.31]& 0.37$\pm$0.09 & 0.28$\pm$0.09 &      & 4.10 & 0.51 & 191 & 2.3 & 66 \\
Sk 18       &  13.6 & 13.13$\pm$0.06 & 16.13 & 0.21$\pm$0.09 & 0.04$\pm$0.03 & 3.64 &      & 0.98 &  55 & 2.9 &[80]\\ 
            & 123.9 & 14.04$\pm$0.03 & 16.38 & 0.41$\pm$0.04 & 0.19$\pm$0.02 &      & 4.32 & 0.78 & 394 & 1.3 & 53 & 5.38 & 83 \\ 
Sk 143\tablenotemark{e} 
            & 132.6 & 14.92$\pm$0.06 &[16.32]& 0.36$\pm$0.11 & 0.25$\pm$0.05 &      & 4.03 & 0.55 & 238 & 0.2 & 45 & 4.15 &  2 \\ 
Sk 155      &     B & 13.07$\pm$0.04 & 16.13 & 0.34$\pm$0.06 & 0.15$\pm$0.04 &      & 4.03 & 0.80 & 131 & 1.9 & 82 & 3.99 &  6 \\ 
\hline
Sk$-$67~5   & 288.3 & 13.62$\pm$0.02 & 16.48 & 0.30$\pm$0.04 & 0.09$\pm$0.02 &      & 3.88 & 0.90 & 133 & 2.8 & 57 & 4.06 &  3 \\ 
            & 295.1 & 13.15$\pm$0.04 & 15.82 & 0.23$\pm$0.09 & 0.09$\pm$0.04 &      & 3.64 & 0.86 &  77 & 1.2 & 57 \\
Sk$-$68~73  &  24.1 & 13.36$\pm$0.04 & 16.31 & 0.19$\pm$0.04 & 0.04$\pm$0.03 & 3.58 &      & 0.97 &  48 & 2.4 &[80]\\ 
            & 289.0 & 13.51$\pm$0.03 & 16.86 & 0.36$\pm$0.05 & 0.23$\pm$0.04 &      & 4.00 & 0.62 & 175 & 9.0 & 57 \\
            & 296.4 & 14.22$\pm$0.03 & 16.77 & 0.47$\pm$0.06 & 0.28$\pm$0.04 &      & 5.11 &[1.00]&2260 &[5.0]& 57 & 4.33 &  9 \\ 
Sk$-$70~115 &  24.2 & 13.25$\pm$0.04 & 16.25 & 0.16$\pm$0.05 & 0.07$\pm$0.03 & 3.38 &      & 0.89 &  30 & 1.9 &[80]\\ 
            &  30.5 & 13.35$\pm$0.03 & 15.69 & 0.24$\pm$0.04 & 0.07$\pm$0.03 & 3.75 &      & 0.92 &  70 & 1.0 &[80]\\ 
            & 220.2 & 13.70$\pm$0.03 & 16.58 & 0.36$\pm$0.06 & 0.16$\pm$0.02 &      & 4.05 & 0.78 & 212 & 4.5 & 53 & 4.48 &  8 \\ 
\hline
MW avg      &       &                &       & 0.21          & 0.07          & 3.60 &      & 0.91 &  50 & 2.7 &[80]&      &    \\
\enddata
\tablecomments{Column densities (cm$^{-2}$) are logarithmic.  Values in square braces are assumed.
The Milky Way average value (last line) corresponds to the ''center of mass'' of JT11's Galactic sample.}
\tablenotetext{a}{Velocities less than 80 (100) km~s$^{-1}$ toward the SMC (LMC) are for Galactic components; $T_{01}$ = 80 K is assumed.  
Group B for Sk~155, with mean velocity $\sim$ 160 km~s$^{-1}$, is defined in Table~\ref{tab:comps}.}
\tablenotetext{b}{Estimated from fits to STIS spectra of O~I, P~II, S~II, and/or Zn~II.}
\tablenotetext{c}{From analysis of C~I fine structure excitation (this paper); ($p/k$)$_{\rm low}$ and $g_{\rm low}$ are the pressure and fractional abundance of the low-pressure gas contributing to each component, assuming that the high-pressure gas has ($f_1$,$f_2$) = (0.38,0.49); $I_{\rm UV}$ is not well constrained for the highest pressure components.}
\tablenotetext{d}{From analysis of H$_2$ excitation; assigned to the velocity nearest that of the observed CH (or strongest Na~I) absorption (Welty et al. 2006, 2013).}
\tablenotetext{e}{Analysis of C$_2$ rotational excitation yields $T$ $\sim$ 25 K, $n_{\rm H}$ $\sim$ 870 cm$^{-3}$, log($p/k$) $\sim$ 4.34 (cm$^{-3}$ K) (Welty et al. 2013).}
\end{deluxetable}

\begin{figure}
\epsscale{0.9}
\plotone{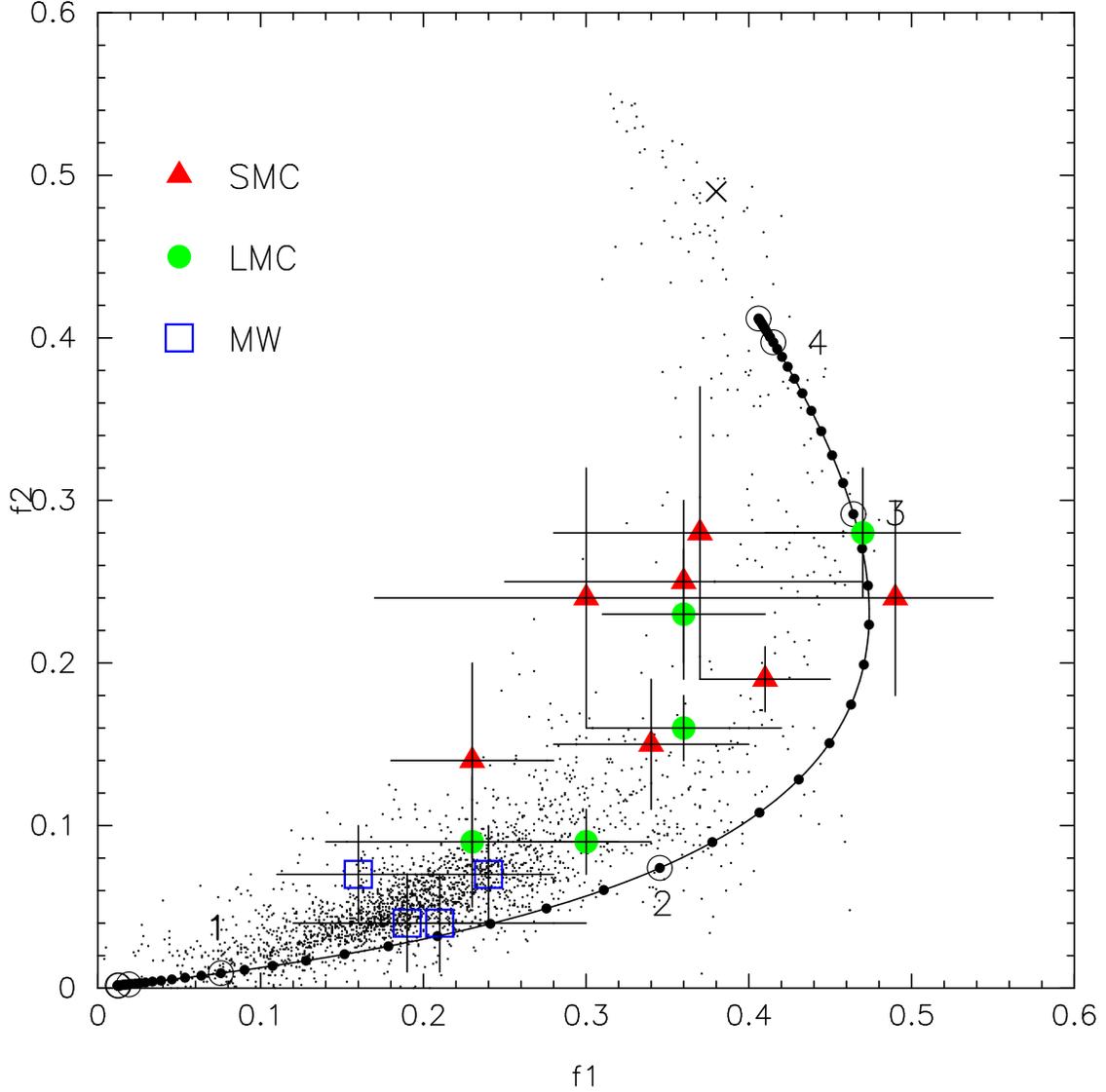}
\caption{Relative fine-structure populations for well characterized components in SMC and LMC sight lines (large colored symbols), compared to Galactic values from JT11 (small black dots).
The curve gives the theoretical populations for $T$ = 80 K and the WJ1 radiation field; the small circles along the curve indicate steps in log($n_{\rm H}$) of 0.1 dex; the larger open circles indicate log($n_{\rm H}$) = 4.0, 3.0, 2.0, 1.0, etc.
The ''x'' at (0.38,0.49) designates the assumed location of the high-pressure component.
While the Galactic components seen toward the Magellanic Clouds (blue open squares) are consistent with the bulk of the JT11 sample, the SMC and LMC components (red triangles and green circles, respectively) generally exhibit higher excitation -- implying higher densities and thermal pressures.}
\label{fig:fsmc}
\end{figure}

\subsection{Excitation of \ion{C}{1}}
\label{sec-c1ex}

Table~\ref{tab:c1ex} lists the relative populations in the two excited \ion{C}{1} fine-structure levels, $f_1$ = $N$(\ion{C}{1}*)/$N$(\ion{C}{1}$_{\rm tot}$) and $f_2$ = $N$(\ion{C}{1}**)/$N$(\ion{C}{1}$_{\rm tot}$), for individual components (or component groups) in our seven sight lines that have well-determined column densities for both the ground and excited levels (from Table~\ref{tab:comps}).
Figure~\ref{fig:fsmc} shows $f_2$ versus $f_1$ for those components, in comparison with the Galactic data from JT11 (for narrow velocity intervals in their sight line sample) and with a theoretical curve (very similar to those shown by JT11) giving the relative populations predicted for hydrogen densities ranging from 0.01 to 10$^5$ cm$^{-3}$, the average Galactic (WJ1) interstellar radiation field (de Boer et al. 1973), and a representative temperature of 80 K (Savage et al. 1977; Tumlinson et al. 2002; Welty et al. 2012).\footnotemark
\footnotetext{In principle, the uncertainties on the column densities of the three fine-structure levels yield a six-sided error box in the ($f_1$,$f_2$) plane (e.g., Jenkins \& Shaya 1979).
The error bars in Fig.~\ref{fig:fsmc} give the approximate dimensions of those error boxes -- and make the plot slightly less complex.}
As discussed by JT11, the curves for higher or lower temperatures follow similar trajectories to that for $T$ = 80 K, with the points corresponding to a given thermal pressure $n_{\rm H}T$ located fairly close to each other (for $n_{\rm H}T$ less than about 10$^5$ cm$^{-3}$ K).
Stronger radiation fields shift the ''origin'' of the curves (at the lowest densities) to higher $f_1$ (and slightly higher $f_2$); see, e.g., Fig.~6 in Jenkins \& Shaya (1979).
The theoretical curve was generated with the code used by Welty (2007), updated to use the more recent determinations of collisional excitation and radiative decay rates employed by JT11.
For $T$ = 30, 80, and 300 K, the curves generated by this code fall slightly below (and/or to the right of) the corresponding curves shown in Fig.~2 of JT11, and the densities corresponding to a given point along each curve are slightly lower -- by 0.18--0.22 dex for the slowly rising part of the curves at lower pressures (0.2 $\la$ $f_1$ $\la$ 0.4) and by 0.10--0.18 dex for the nearly vertical part of the curves at higher pressures (0.1 $\la$ $f_2$ $\la$ 0.4).\footnotemark
\footnotetext{Roughly 0.06 dex of the difference may be due to our neglect of collisions with H$_2$ (which should be reasonable for most of the components in our small sample), whereas the curves in JT11 assume $f$(H$_2$) = 0.6 (E. Jenkins 2015, private communication).}
Those offsets were confirmed by applying our iterative analysis (see below) to a small set of ''components'' in JT11's Galactic sample, covering the ranges 30 $\la$ $T$ $\la$ 110 K and 3.0 $\la$ log($p/k$) $\la$ 4.4 (cm$^{-3}$ K).
In order to facilitate comparisons with the results of JT11, the densities and pressures derived below are adjusted for those offsets.

Most of the points in JT11's Galactic sample have $f_1$ $\la$ 0.3 and $f_2$ $\la$ 0.1, and the ''center of mass'' of all the Galactic points is at about ($f_1$,$f_2$) = (0.21,0.07) -- slightly above the theoretical curves for ''reasonable'' temperatures for cold, neutral gas.
The points falling above and to the left of the theoretical curves are thought to be due to contributions from mixtures of low- and high-pressure gas -- dominated in most cases by the lower pressure gas.
The four Galactic components in Table~\ref{tab:c1ex} (those with $v$ $\la$ 80 km~s$^{-1}$ toward the SMC or $v$ $\la$ 100 km~s$^{-1}$ toward the LMC) have relative fine-structure populations quite consistent with the majority of those in that larger Galactic sample.
As might have been surmised by inspection of the line profiles (Appendix Figures~\ref{fig:sk13c1} through \ref{fig:sk70d115c1}), however, the SMC and LMC components exhibit systematically higher excitation than most of the Galactic points -- with $f_1$ $\ge$ 0.3 and/or $f_2$ $\ge$ 0.09 in all cases -- indicating both systematically higher densities (or thermal pressures) in those components and generally higher fractions of very high pressure gas.
[The temperatures for interstellar clouds, as determined from H$_2$, are similar (on average) in the Milky Way, LMC, and SMC (Tumlinson et al. 2002; Welty et al. 2012), so that such comparisons can be made in terms of either density or thermal pressure.]
While two of the Magellanic Clouds points (Sk~13/$v$=147.6, Sk$-$68~73/$v$=296.4) are fairly close to the theoretical curve (consistent with gas at a single pressure), most fall well above and to the left of the curve -- within the high-excitation tail of the Galactic distribution -- suggesting that they represent mixtures of gas at different pressures.

\subsection{Estimation of Thermal Pressures}
\label{sec-pres}

The excitation of \ion{C}{1} in a given sight line depends both on the temperature and density in the gas and on the strength of the ambient UV radiation field -- though the effects of the UV field are most significant when the densities are relatively low.
If the temperature and UV field are known, then values for the pressure/density and for the fractions of low- and high-pressure gas ($g_{\rm low}$ and $g_{\rm high}$, respectively) can be obtained by comparing the observed ($f_1$,$f_2$) with the appropriate theoretical curve, with the additional assumption that the point representing ($p/k$)$_{\rm high}$ is at ($f_1$,$f_2$) = (0.38,0.49) (Fig.~\ref{fig:fsmc}; as in JT11).
For observed points falling on the theoretical curve, $g_{\rm low}$ = 1.0, and ($p/k$)$_{\rm low}$ (and thus the local density) can be read directly from the curve.
For points in the ''allowed'' region above and to the left of the theoretical curve, the line between the assumed high-pressure point and the observed point is extended downward to intersect with the equilibrium curve (for the given $T$ and UV field) to obtain values for ($p/k$)$_{\rm low}$, and $g_{\rm low}$ and $g_{\rm high}$ are determined from the relative distances between the observed point and the two endpoints (see Fig.~4 in JT11).

While the strength of the UV field can be estimated for all seven of our SMC and LMC sight lines from the excitation of H$_2$ (e.g., Welty et al. 2006), those values depend on the column densities in the $J$=4,5 rotational levels, which can be rather uncertain (e.g., Tumlinson et al. 2002).
Following JT11, an iterative procedure was therefore employed to obtain estimates for both the density (and thus the thermal pressure) and the strength of the UV field from the observed excitation of \ion{C}{1} in each component, assuming that the temperature obtained from the populations in the lowest two rotational levels of H$_2$ characterizes the gas traced by \ion{C}{1} (for all LMC or SMC components in the sight line; 80 K was assumed for any Galactic components).
In each iteration, the density is first estimated from the \ion{C}{1} excitation (as just described), for a given UV field strength (represented as $I_{\rm UV}$ times the fiducial WJ1 field, with $I_{\rm UV}$ initially set to 1.0).
Next, that density is used to obtain a new estimate for $I_{\rm UV}$ from the ionization equilibrium of carbon, using an estimate for the column density of singly ionized carbon in that component (column 4 of Table~\ref{tab:c1ex}) derived from observations of other little-depleted dominant ions (e.g., \ion{O}{1}, \ion{P}{2}, \ion{S}{2}, and/or \ion{Zn}{2}) via fits to lines from those species in the STIS spectra of our seven targets.\footnotemark
\footnotetext{For the SMC components toward Sk~13 and Sk~143, estimates of $N$(C~II) could not be obtained from fits to lines of other little-depleted species, due to a combination of spectral coverage, low S/N, and/or strongly saturated lines for the available species.
For the four components toward Sk~13, $N$(C~II)/$N$(C~I) = 10$^3$ was assumed -- similar to the values found for other SMC and LMC components (and for JT11's Galactic sample) and consistent with the total sight line $N$(C~II) estimated from $N$(H$_{\rm tot}$) and a depletion of $-$0.2 dex.
For the main component toward Sk~143, log[$N$(C~II)] = 16.32 (a factor $\sim$2 less than the value estimated for the total sight line) was assumed.
As noted in the text, the uncertainties in these $N$(C~II) affect primarily the derived $I_{\rm UV}$ -- but not the pressures/densities.} 
Contributions to the electron density from both photoionization of heavy elements and cosmic-ray ionization of hydrogen, and both radiative and grain-assisted recombination (Weingartner \& Draine 2001), are included in the calculation of ionization equilibrium.
The heavy element abundances, cosmic-ray ionization rate, and grain-assisted recombination rates are scaled by factors of 0.5 and 0.2 for the LMC and SMC, respectively, to account (approximately) for metallicity-related effects.\footnotemark
\footnotetext{See Abdo et al. (2010) and Ackermann et al. (2016) for observational constraints on the cosmic-ray fluxes in the SMC and LMC, respectively.}
The new value for $I_{\rm UV}$ is then used to generate a revised theoretical curve for an updated determination of the pressure/density.
Convergence to a solution for the pressure/density and UV field is generally achieved within about 10--20 iterations.
The final pressure/density values are then adjusted for the small offsets found in comparisons with the results of JT11 (which also suggest that the derived $g_{\rm low}$ and $I_{\rm UV}$ listed in Table~\ref{tab:c1ex} may be slightly underestimated).
Inclusion of cosmic-ray ionization of H significantly increases the inferred electron densities and UV field strengths, but has only a minor effect on the derived pressures/densities.
Uncertainties in the ionization equilibrium calculations (e.g., Weingartner \& Draine 2001; Welty et al. 2003) -- including those from the estimation of $N$(\ion{C}{2}) -- and inclusion of dielectronic recombination (which may be significant for \ion{C}{1} at low temperatures; Badnell 2006) would also affect primarily the derived UV field strengths, with little impact on the derived pressures/densities.

As expected from the observed relative populations in the excited \ion{C}{1} fine structure states (Fig.~\ref{fig:fsmc}), the four Milky Way components included in Table~\ref{tab:c1ex} (toward Sk~18, Sk$-$68~73, and Sk$-$70~115) all have thermal pressures similar to typical Galactic values (JT11) -- with log($p/k$)$_{\rm low}$ $\sim$ 3.38--3.75 (cm$^{-3}$ K) and $g_{\rm low}$ $\sim$ 0.9--1.0; the corresponding local densities $n_{\rm H}$ range from about 30 to 70 cm$^{-3}$, for the assumed $T$ = 80 K.
The 12 SMC and LMC components in Table~\ref{tab:c1ex}, however, tend to be characterized by higher thermal pressures [log($p/k$)$_{\rm low}$ $\sim$ 3.56--5.11 (cm$^{-3}$ K)] and/or higher fractions of high-pressure gas ($g_{\rm low}$ $\sim$ 0.5--1.0); the corresponding densities, for the low-pressure component, range from about 55 to 2260 cm$^{-3}$.
The derived UV field strengths are generally lower (1 $\la$ $I_{\rm UV}$ $\la$ 5) than those estimated from H$_2$ excitation -- particularly for Sk~13 and Sk~18 -- and, for these higher pressures, have relatively minor effects on the excitation.
These general results do not appear to be biased by the requirement that all three fine-structure levels be securely detected in those components, as the SMC and LMC components in which the typically weaker lines from \ion{C}{1}** are not detected generally do exhibit higher $N$(\ion{C}{1}*)/$N$(\ion{C}{1}) ratios than the values found for most of the ''components'' in JT11's Galactic sample -- consistent with higher pressures and densities in the Magellanic Clouds.
(Likewise, there is no apparent trend of higher excitation for the higher column density components in JT11's Galactic \ion{C}{1} data.)
Unfortunately, the upper limits on $N$(\ion{C}{1}**) in the lower column density SMC and LMC components generally cannot distinguish between the typical Galactic and (higher) Magellanic Clouds pressures.

\section{DISCUSSION}
\label{sec-disc}

\subsection{Thermal Pressures in the SMC and LMC}
\label{sec-presmc}

The values of ($p/k$)$_{\rm low}$ and $g_{\rm low}$ listed in Table~\ref{tab:c1ex} are not the first indications of higher thermal pressures for cool, predominantly neutral diffuse atomic and molecular clouds in the ISM of the Magellanic Clouds:
\begin{itemize}
\item{Observations of \ion{C}{1} absorption in lower resolution spectra of SN~1987A (LMC), AV~95 (SMC), and Sk$-$69~246 (LMC) had suggested somewhat higher population of the excited fine-structure levels there (Welty et al. 1999a; Andr\'{e} et al. 2004) -- but both the absolute column densities and the distribution of the absorption among the various individual components in those sight lines (unresolved in the UV spectra) are somewhat uncertain.}
\item{Observations of H$_2$ absorption toward SMC and LMC targets have revealed that the relative populations in the higher H$_2$ rotational levels (e.g., $J$=4,5) generally are higher in the Magellanic Clouds than in typical Galactic sight lines (Tumlinson et al. 2002; Andr\'{e} et al. 2004; Cartledge et al. 2005; Xue et al., in preparation).
In principle, those levels can be populated radiatively, collisionally (e.g., in shocks), or in H$_2$ formation (e.g., Shull \& Beckwith 1982).
If photon pumping is primarily responsible for populating the higher $J$ levels (e.g., Jura 1974, 1975), then estimates for both the local hydrogen density $n_{\rm H}$ and the strength of the ambient UV radiation field $I_{\rm UV}$ may be obtained from expressions proportional to $N$($J$=4,5); $n_{\rm H}$ also depends on the assumed rate coefficient $R$ for H$_2$ formation (e.g., Lee et al. 2002; Welty et al. 2006).
As the temperatures derived for gas in the Magellanic Clouds (from the lowest two H$_2$ levels) are very similar to those found for the Galactic ISM, the higher $J$=4,5 populations in the LMC and SMC suggest that both the density and the thermal pressure are generally higher there.
The uncertainties on $N$($J$=4,5) can be large, however, as the lines from those levels are often on the flat part of the curve of growth (Tumlinson et al. 2002), and (again) the multiple individual components present in most sight lines cannot be distinguished in the lower-resolution {\it FUSE} spectra (Welty et al. 2006; Welty \& Crowther, in preparation).
For the small sample in Table~\ref{tab:c1ex}, the pressures estimated from H$_2$ generally agree with those determined from \ion{C}{1} to within factors of about 2.5; the values inferred from H$_2$ are much higher toward Sk~13 and Sk~18, however.}
\item{Analysis of the C$_2$ rotational excitation in the sight line toward Sk~143 (SMC) has yielded values for $n_{\rm H}$ and $p/k$ that are 2--4 times higher than those found in most Galactic sight lines in which C$_2$ is detected (Sonnentrucker et al. 2007; Welty et al. 2013 and in preparation).
The pressure inferred from C$_2$ is slightly higher than the values obtained from \ion{C}{1} and H$_2$ (Table~\ref{tab:c1ex}), consistent with C$_2$ tracing somewhat denser gas.
The densities and pressures obtained from C$_2$ scale with the strength of the radiation field in the near-IR, however, which is often not well constrained.}
\end{itemize}
The pressures and densities inferred here from STIS echelle spectra of \ion{C}{1} should be the most reliable values currently available for cool, predominantly neutral diffuse gas in the Magellanic Clouds -- as they are based on the highest resolution UV spectra and are relatively insensitive to both the typically rather uncertain radiation field (in the UV and/or near-IR) and other assumed parameters.

The higher thermal pressures found for cool, predominantly neutral gas in the Magellanic Clouds are qualitatively consistent with the predictions of models for systems characterized by lower metallicities (with correspondingly lower dust-to-gas ratios) and stronger UV fields (e.g., Wolfire et al. 1995).
In lower metallicity systems, reduced cooling, due to lower abundances of the usual gas-phase coolants, is at least initially balanced by lower photoelectric heating (less dust) -- as both processes scale (at least roughly) with the metallicity.
When the photoelectric heating becomes less than the heating from X-rays, however, the equilibrium temperature and pressure required for stable cold, neutral clouds are increased.
For metallicities 0.3 and 0.1 times Solar (and dust-to-gas ratios correspondingly smaller than those in the local Galactic ISM), for example, the models predict that the minimum pressures would be increased by factors of about 1.3 and 1.8, respectively (Fig.~6b in Wolfire et al. 1995).
Similarly, higher densities (and pressures) are required for the gas-phase coolants to balance the increased heating in clouds subject to stronger ambient radiation fields.
For radiation fields enhanced by factors of 3, 10, and 100, the minimum pressures would be increased by factors of about 2.2, 5.2, and 25, respectively (Fig.~7 in Wolfire et al. 1995).
While a dependence on metallicity might suggest that, on average, slightly higher pressures should be found for the SMC, there is no obvious systematic difference between the pressures derived for the LMC and SMC sight lines in this small sample; local environmental factors (including enhanced radiation fields) may contribute to the higher pressures in specific cases.

In the Galactic ISM, higher thermal pressures do tend to be found where the UV radiation field is enhanced (e.g., Fig.~8 in JT11) -- with perhaps a slightly steeper dependence than that predicted by the models of Wolfire et al. (1995).
In our Magellanic Clouds sample, some of the lowest pressures [log($p/k$)$_{\rm low}$ $\sim$ 3.64--3.88 (cm$^{-3}$ K)] -- and some of the lowest fractions of very high pressure gas -- are found toward Sk$-$67~5, which also has fairly low values for $I_{\rm UV}$ ($\sim$ 1--3, as inferred from both \ion{C}{1} and the high-$J$ H$_2$ populations).
On the other hand, the sight line toward Sk~143 has even lower $I_{\rm UV}$, but both higher pressure [log($p/k$)$_{\rm low}$ $\sim$ 4.03 (cm$^{-3}$ K)] and a higher fraction of very high pressure gas.
The sight line toward Sk~13 may have somewhat higher $I_{\rm UV}$ (from H$_2$) and one component with high pressure, but also has several components with relatively low ($p/k$)$_{\rm low}$.
The sight line toward Sk~18 has the highest $I_{\rm UV}$ from H$_2$ ($\sim$ 80), but not the highest pressure.
While there is thus no clear relationship between the pressure and the ambient UV radiation field in this small sample, the values for $I_{\rm UV}$, derived either from the higher-$J$ H$_2$ populations or from carbon ionization equilibrium, can be rather uncertain (e.g., Tumlinson et al. 2002).

As all of our Magellanic Clouds targets are OB stars, all of the sight lines will contain at least some ionized gas -- which, in principle, might contribute to the observed abundances and excitation of \ion{C}{1} (Jenkins \& Shaya 1979). 
Using fluxes from model stellar atmospheres, Jenkins \& Tripp (2001) calculated that \ion{C}{1} column densities up to about 3 $\times$ 10$^{13}$ cm$^{-2}$ might be seen from the \ion{H}{2} regions of Galactic OB stars.
As noted in Table~\ref{tab:env}, Sk~143 (O9.7~Ib), Sk$-$67~5 (O9.7~Ib), Sk$-$68~73 (Of/WN), and Sk$-$70~115 (O6.5~Iaf) appear to be associated with distinct \ion{H}{2} regions; Sk~13 (B2~Ia) and Sk~18 (O7~III) probably contribute to the ionization in the N19 complex; only Sk~155 (O9~Ib) appears to have no associated nebulosity (though some ionized gas is present).
Absorption from \ion{Al}{3}, a tracer of ionized gas, is seen at velocities overlapping the main \ion{C}{1} components (and generally over a somewhat wider velocity range) in all seven sight lines.
While strong absorption from \ion{Si}{2}* (which can be present in both neutral and ionized gas) is seen at the same velocities as \ion{C}{1} toward Sk~13 and Sk$-$68~73, fairly strong absorption from \ion{O}{1}* and \ion{O}{1}** (which trace predominantly neutral gas) is also present -- and the ratio $N$(\ion{O}{1}*)/$N$(\ion{O}{1}**), of order 1.4$\pm$0.2 toward Sk$-$68~73, implies that $T$ $\la$ 330 K (e.g., Fig.~15 in JT11).
Moderately strong \ion{Si}{2}* and weaker \ion{O}{1}* are present toward Sk$-$70~115, but not at the velocity of the strongest \ion{C}{1}; weaker \ion{Si}{2}* is seen toward Sk~18, Sk~143, Sk~155, and Sk$-$67~5.
It thus seems unlikely that ionized regions contribute significantly to the \ion{C}{1} observed in these LMC and SMC sight lines -- or to the higher thermal pressures derived from its excitation -- especially given the much lower carbon abundances in the Magellanic Clouds.

Some of the higher thermal pressures in these Magellanic Clouds sight lines, however, may be related to energetic activity (e.g., due to stellar winds, star formation, and/or supernova remnants) in the particular regions where the clouds are located.
For example, the three sight lines in which the highest thermal pressures are found all appear to probe complex regions containing both ionized and molecular gas and showing indications of such energetic activity (see, e.g., the optical and near-IR images for these sight lines presented on the {\it FUSE} Magellanic Clouds Legacy Project website; Blair et al. 2009).
Both Sk~13 and Sk~18 are located in the southwestern part of the SMC ''bar'', in the N19 complex -- which includes a number of \ion{H}{2} regions and molecular clumps and several supernova remnants, producing complex internal kinematics (e.g., Chu \& Kennicutt 1988; Rubio et al. 1993a; Rosado et al. 1994; Dickey et al. 2001); the values of log($p/k$)$_{\rm low}$ in the main components in these two sight lines are 4.82 and 4.32 (cm$^{-3}$ K), respectively.
Even higher thermal pressure [log($p/k$)$_{\rm low}$ $\sim$ 5.11 (cm$^{-3}$ K)] is found for the main component toward Sk$-$68~73, located in the central part of the LMC in the N44 complex -- which also contains a number of ionized and molecular regions and which shows evidence for fairly recent supernovae (e.g., Meaburn \& Laspias 1991; Kim et al. 1998; Naz\'{e} et al. 2002; Chen et al. 2009; Wong et al. 2011).
Somewhat lower thermal pressures [log($p/k$)$_{\rm low}$ $\sim$ 3.64--4.03 (cm$^{-3}$ K)] are found for the simpler, more isolated, more quiescent regions toward Sk$-$67~5 and Sk~143, and also toward Sk~155, where much of the gas seen in absorption apparently is foreground to the ionized and molecular gas seen in emission (Welty et al. 2012 and in preparation).
Even those sight lines, however, exhibit higher thermal pressures and/or higher fractions of very high pressure gas than those found in most Galactic sight lines in the JT11 sample.
The current sample of Magellanic Clouds sight lines is still very small, however, and may not be entirely representative of those two lower metallicity galaxies.

\begin{figure}
\epsscale{0.9}
\plotone{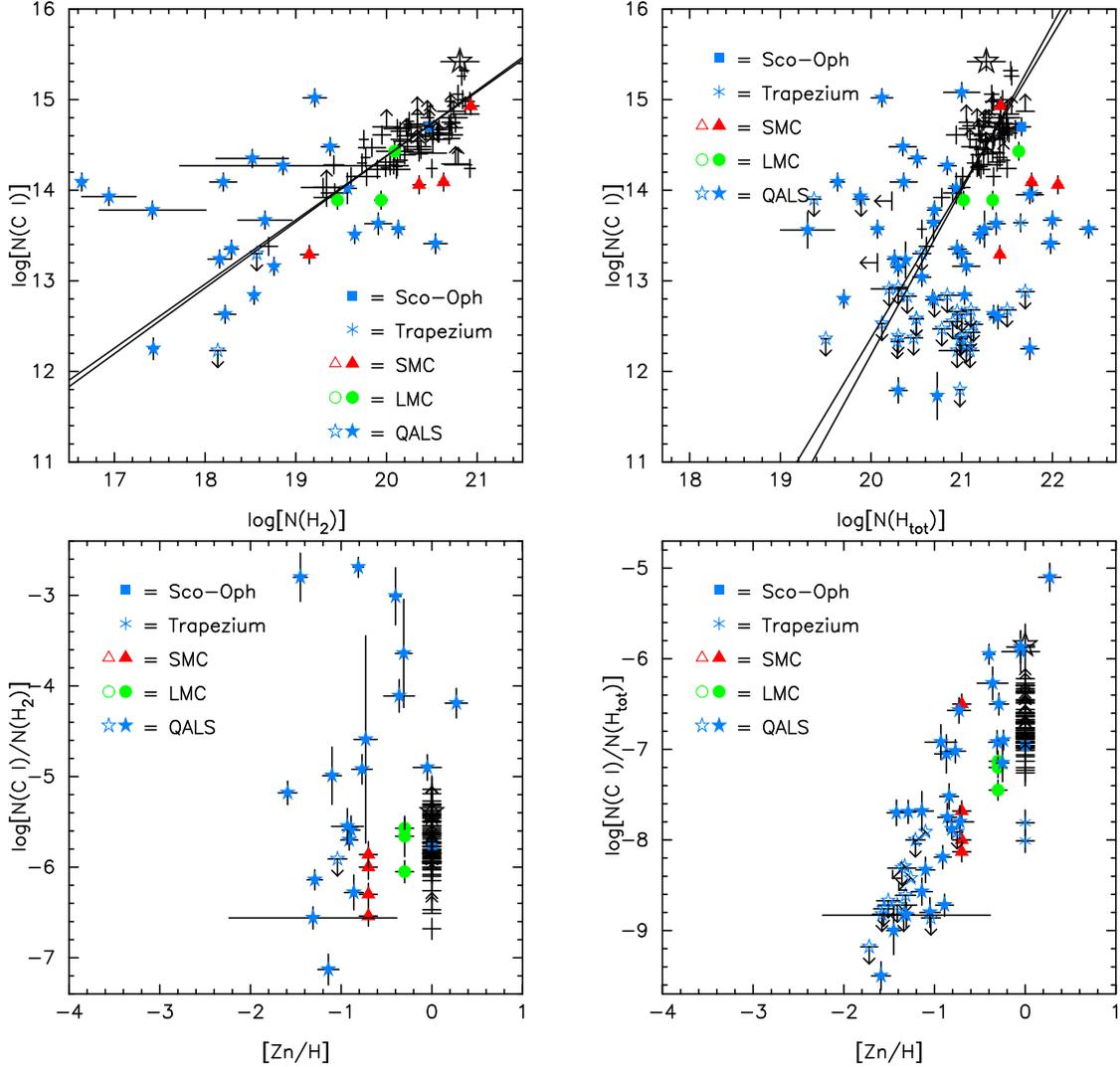}
\caption{Column density of \ion{C}{1}$_{\rm tot}$ versus column densities of H$_2$ ({\it upper left}) and H$_{\rm tot}$ ({\it upper right}) and the corresponding ratios of those quantities versus the metallicity indicator [Zn/H] ({\it lower left , lower right}), for QSO absorbers (stars) and for sight lines in the Milky Way (plain crosses) and Magellanic Clouds.
The solid lines in the upper panels are fits to the Galactic data only, as in Fig.~\ref{fig:c1abund}.
NOTE:  The $N$(\ion{C}{1}$_{\rm tot}$) values for the Galactic and Magellanic Clouds sight lines were obtained using the $f$-values derived by JT11, while the values for the QSO absorbers were obtained using $f$-values tabulated by Morton (2003) -- so the two sets of data are NOT precisely comparable (see Sec.~\ref{sec-c1f}).
The two sets are plotted together here primarily to compare the QSO absorber values with trends seen in the Galactic and Magellanic Clouds sight lines.
Much of the scatter in the QSO absorber values seen in the two upper panels may be due to differences in metallicity among the various systems -- as plots of the ratios in the lower two panels reveal similar, generally increasing trends with [Zn/H] for the absorbers, the Magellanic Clouds, and our Galaxy.}
\label{fig:c1qso}
\end{figure}

\subsection{\ion{C}{1} in QSO Absorption-line Systems}
\label{sec-qso}

These results regarding the abundance and excitation of \ion{C}{1} in the LMC and SMC may also inform our understanding of the \ion{C}{1} absorption detected in some quasar absorption-line systems -- many of which are also characterized by sub-Solar metallicities.
While most QSO absorbers appear to be due to relatively warm, low density gas, with very low molecular fractions and fairly mild depletions, the (relatively few) systems in which \ion{C}{1} is detected generally contain colder, denser gas, with higher dust and molecular content (e.g., Srianand et al. 2005; Ledoux et al. 2015).
Most of those systems are either confirmed or likely damped Lyman-$\alpha$ systems [DLAs; with $N$(H) $\ge$ 2 $\times$ 10$^{20}$ cm$^{-2}$] or fairly strong sub-DLAs.
Measurements of the excitation of \ion{C}{1} in those systems enable limits to be placed on both the local physical conditions in more distant galaxies and the redshift dependence of the temperature of the CMB (e.g., Songaila et al. 1994; Ge et al. 1997; Quast et al. 2002).
Ledoux et al. (2015) have measured the equivalent widths of the $\lambda$1560 and $\lambda$1656 \ion{C}{1} multiplets for 66 systems found in a sample of low-resolution ($R$ $\sim$ 2000) SDSS spectra of 41,696 quasars -- but higher resolution spectra will be needed to determine accurate column densities and to assess the fine-structure excitation in those systems.
Moderately high resolution spectra (mostly from Keck/HIRES and/or VLT/UVES) have yielded detections of \ion{C}{1} in a smaller number of systems, and fits to the line profiles have yielded column densities for the ground and excited fine-structure states for the ''individual'' components discernible in the spectra (Quast et al. 2002; Srianand et al. 2005, 2008; Noterdaeme et al. 2007a, 2010, 2015; Jorgenson et al. 2010; Carswell et al. 2011; Ma et al. 2015).
Srianand et al. (2005) performed a simplified analysis of the \ion{C}{1} excitation for components in eight systems (considering only collisions with atomic hydrogen and generally assuming $T$ = 100 K) and found somewhat higher pressures than those typically found in the local Galactic ISM.
Jorgenson et al. (2010) used an approach more similar to that employed by JT11 (assuming ionization equilibrium for carbon, but also solving for $T$).
For 11 components in six systems, Jorgenson et al. found median values for $n_{\rm H}$ $\sim$ 70 cm$^{-3}$, $T$ $\sim$ 50 K, and log($p/k$) $\sim$ 3.86 (cm$^{-3}$ K).
That median pressure is nearly a factor of 2 higher than the JT11 Galactic ''center of mass'' value and also is higher than the values estimated for those systems from the excitation of \ion{C}{2}.
Jorgenson et al. concluded that the \ion{C}{1} generally traces a cold, neutral component in those systems, perhaps in denser clumps than the rest of the neutral material.

\begin{figure}
\epsscale{0.9}
\plotone{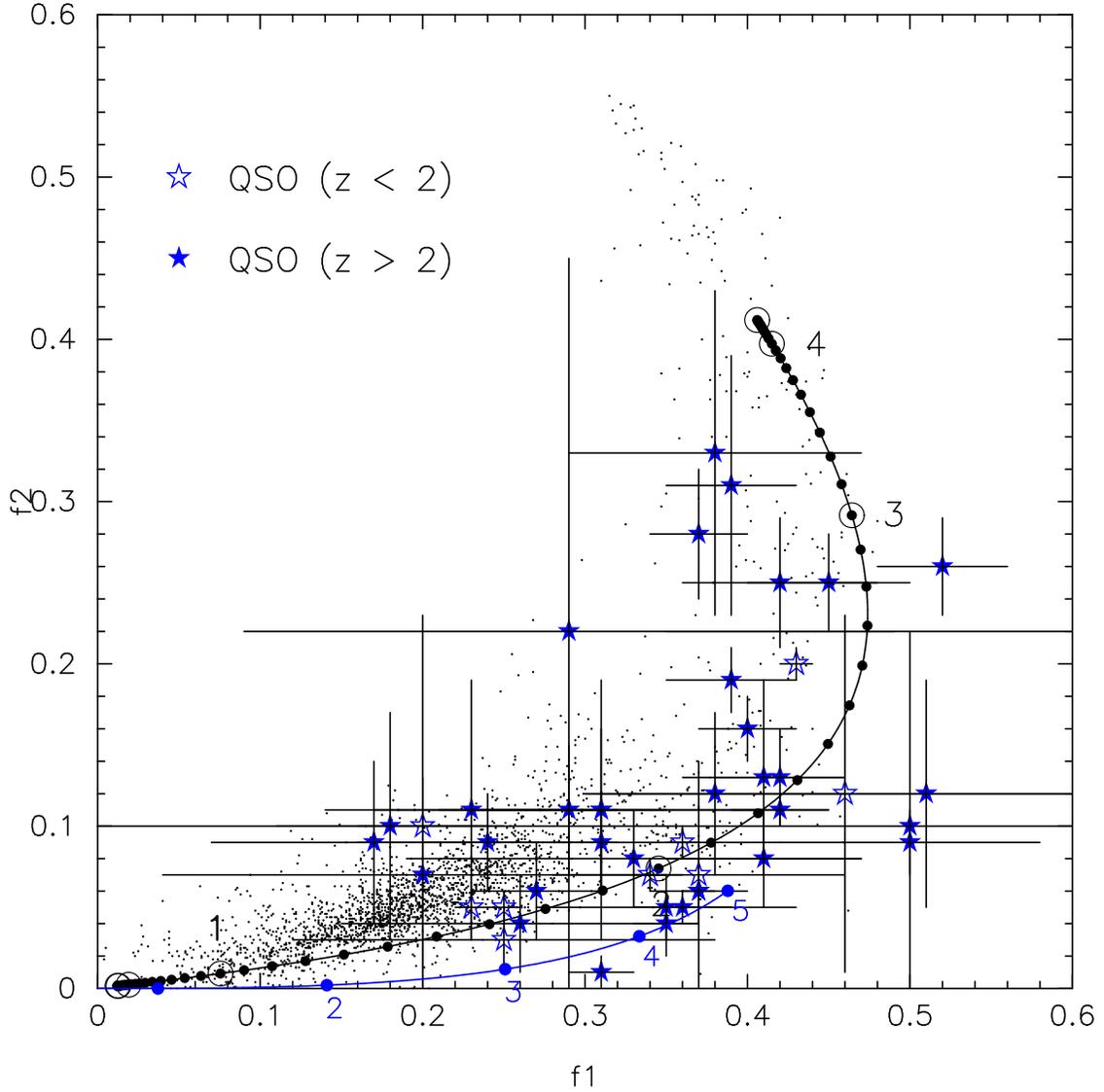}
\caption{Relative fine-structure populations for well characterized components in QSO absorption-line systems (blue stars), compared to Galactic values from JT11 (small black dots).
The solid black curve gives the theoretical populations for $T$ = 80 K and the WJ1 radiation field; the small circles along the curve indicate steps in log($n_{\rm H}$) of 0.1 dex; the larger open circles indicate log($n_{\rm H}$) = 4.0, 3.0, 2.0, 1.0, etc.
The lower solid blue curve gives the populations in equilibrium with the Cosmic Microwave Background, for $z$ = 0 to 5 -- representing the minimum ($f_1$,$f_2$) that would be found for a given $z$; the solid circles along the curve indicate $z$ = 5, 4, 3, 2, 1.
Some of the QSO absorber components are consistent with the bulk of the Galactic sample, but most exhibit somewhat higher excitation.
As all the QSO absorbers here are at $z$ $<$ 3.1, the CMB generally plays a fairly minor role in the excitation.}
\label{fig:fsqso}
\end{figure}

Figure~\ref{fig:c1qso} compares the column densities of \ion{C}{1}$_{\rm tot}$ with those of H$_2$ and H$_{\rm tot}$, for QSO absorbers (Appendix Table~\ref{tab:c1qso}) and for sight lines in the Milky Way and Magellanic Clouds (from Tables~\ref{tab:gal} and \ref{tab:mc}).
While the two sets of data are not precisely comparable, due to differences in the assumed \ion{C}{1} $f$-values (Sec.~\ref{sec-apqso}), they are plotted together in order to compare general trends and scatter.
In the two upper panels, there appears to be much more scatter in the QSO absorber values, so that general trends -- like those seen for the Galactic and Magellenic Clouds sight lines -- are difficult to identify (see also Srianand et al. 2005).
Much of that scatter, however, may be due to differences in metallicity -- as seen in comparisons of Galactic and Magellanic Clouds sight lines (Fig.~\ref{fig:c1abund}); the absorbers with higher (lower) $N$(\ion{C}{1}$_{\rm tot}$), for a given $N$(H$_2$) or $N$(H$_{\rm tot}$), tend also to be characterized by higher (lower) metallicities (as estimated from [Zn/H]\footnotemark).
Some of the scatter could also be due to different amounts of warmer, more diffuse neutral gas -- containing H but little \ion{C}{1} -- in the different systems.
\footnotetext{[Zn/H] = log[$N$(\ion{Zn}{2})/$N$(H$_{\rm tot}$)] $-$ log(Zn/H)$_{\odot}$, where the Solar zinc abundance log(Zn/H)$_{\odot}$ = $-$7.37 dex (Lodders 2003).}
Moreover, a good correlation between $N$(\ion{C}{1}$_{\rm tot}$) and $N$(H$_2$) is not expected for $N$(H$_2$) $\la$ 10$^{18}$ cm$^{-2}$, where the H$_2$ is not fully self-shielded; see, e.g., the analogous plots of $N$(\ion{Na}{1}) and $N$(\ion{K}{1}) versus $N$(H$_2$) in Welty \& Hobbs (2001).
Plots of the ratios $N$(\ion{C}{1}$_{\rm tot}$)/$N$(H$_2$) and $N$(\ion{C}{1}$_{\rm tot}$)/$N$(H$_{\rm tot}$) versus metallicity (the two lower panels in the figure) reveal similar trends (higher ratios for higher metallicities) for the QSO absorbers, the Magellanic Clouds, and our Galaxy [when the four absorbers with highest $N$(\ion{C}{1}$_{\rm tot}$)/$N$(H$_2$), which all have $N$(H$_2$) $<$ 10$^{18}$ cm$^{-2}$, are removed]. 
For the QSO absorbers, the fairly steep relationship between $N$(\ion{C}{1}$_{\rm tot}$)/$N$(H$_{\rm tot}$) and metallicity ({\it lower right}) should be similar to the relationship between $N$(\ion{C}{1}$_{\rm tot}$) and $N$(\ion{C}{2}); the slope ($\sim$ 2) is then consistent with expectations from considerations of ionization equilibrium -- just as for $N$(\ion{C}{1}$_{\rm tot}$), $N$(\ion{Na}{1}), and $N$(\ion{K}{1}) versus $N$(H$_{\rm tot}$) in the Galactic ISM (Sec.~\ref{sec-c1ab} above; Welty \& Hobbs 2001).
Much of the remaining scatter in $N$(\ion{C}{1}$_{\rm tot}$)/$N$(H$_{\rm tot}$), for a given [Zn/H], is likely due to differences in the strength and shape of the local radiation field (as modified by the UV extinction) -- as appears to be the case for the Galactic points at [Zn/H] = 0.0.

The top section of Appendix Table~\ref{tab:c1qso} lists both total \ion{C}{1} column densities and the relative populations of the excited levels (which should be less sensitive to differences in adopted $f$-values), for 46 ''individual'' components (in 19 systems) in which all three \ion{C}{1} levels have been detected, as well as system-wide values for the column densities of H, H$_2$, and \ion{C}{1}$_{\rm tot}$ and the metallicity indicator [Zn/H].
Those systems range in redshift from about 1.1 to 3.3, with total hydrogen column densities from about 0.8 to 250 $\times$ 10$^{20}$ cm$^{-2}$, molecular fractions $f$(H$_2$) from less than 0.01 to nearly 0.3, and metallicities (as gauged from [Zn/H]) from about $-$1.9 to +0.3 dex.
Because many of the QSO absorbers have metallicities well below Solar and relatively mild depletions of refractory elements (e.g., from [Fe/Zn]), and because of the expected increase in $T_{\rm CMB}$ at higher redshifts, those absorbers should be characterized by somewhat higher thermal pressures -- and should exhibit corresponding higher excitation of \ion{C}{1} -- than those found for most clouds in the local Galactic ISM (if the local radiation fields are comparable).
Figure~\ref{fig:fsqso} compares the \ion{C}{1} excitation in those QSO absorber components with the Galactic values determined by JT11.
As in Fig.~\ref{fig:fsmc}, the solid black line shows the relative populations predicted for $T$ = 80 K, the standard Galactic interstellar radiation field, and densities ranging from 0.01 to 10$^5$ cm$^{-3}$.
The shorter solid blue line shows the populations that would be obtained if the CMB were the only source of excitation, with $T_{\rm CMB}$ = 2.725 (1 + $z$) K, for redshifts 0 $\le$ $z$ $\le$ 5 -- i.e., giving the minimum ($f_1$,$f_2$) that would be found for each $z$ (e.g., Fig.~12 in Jorgenson et al. 2010).
While about one quarter of the QSO absorber components exhibit excitation consistent with the bulk of the Galactic \ion{C}{1} sample, the rest have higher values for $f_1$ ($>$ 0.28) and (in many cases) for $f_2$ ($>$ 0.10).
The majority of the QSO absorber components thus do appear to be characterized by somewhat higher \ion{C}{1} excitation (and thus higher thermal pressures), as found for the main LMC and SMC components in our small sample (Fig.~\ref{fig:fsmc}, Table~\ref{tab:c1ex}); similar results have been reported by Srianand et al. (2005) and Jorgenson et al. (2010).
The relative populations in the second excited state (\ion{C}{1}**) are smaller, on average, for the QSO absorbers than for the Magellanic Clouds sight lines, however, suggesting that the fraction of gas at very high pressures may generally be smaller in the QSO absorbers.
There are no obvious trends of increasing $f_1$ with either decreasing metallicity or increasing $z$, but the sample is still very small.
In most cases, the CMB does not dominate the excitation of \ion{C}{1}, as most of the components in Table~\ref{tab:c1qso} (as well as others in which \ion{C}{1}** is not detected) have $T_{01}$ (for \ion{C}{1}) appreciably greater than $T_{\rm CMB}$($z$), and as all the absorbers are at $z$ $<$3.3 (for which the predicted minimum $f_1$ and $f_2$ are low; see also Quast et al. 2002; Srianand et al. 2005; Jorgenson et al. 2010).
At $z$ = 3, the calculations of Silva \& Viegas (2002) indicate that the excitation to \ion{C}{1}* due to the CMB exceeds that due to collisions only for densities less than about 40 cm$^{-3}$ (at $T$ = 100 K).
In general, analysis of the fine-structure excitation of \ion{C}{1} can yield only upper limits on $T_{\rm CMB}(z)$ in quasar absorption-line systems.

\section{SUMMARY / CONCLUSIONS}
\label{sec-sum}

We have examined the abundance and fine-structure excitation of \ion{C}{1} for seven diverse sight lines in the Magellanic Clouds, based on analyses of high-resolution UV spectra obtained with {\it HST}/STIS.
These are the only Magellanic Clouds sight lines for which high-resolution UV spectra covering a number of \ion{C}{1} multiplets have been obtained.
Detailed fits were made to the absorption-line profiles of seven \ion{C}{1} multiplets, guided by fits to the corresponding profiles of the interstellar \ion{Na}{1} lines in higher resolution and/or higher S/N ratio optical spectra.
The combination of high spectral resolution (for both optical and UV data) and coverage of both weak and strong lines (for both \ion{Na}{1} and \ion{C}{1}) enabled the determination of well constrained column densities for the ground and excited fine-structure states of \ion{C}{1} in all seven sight lines.

Comparisons based on a new compilation of column densities of H, H$_2$, \ion{Na}{1}, \ion{K}{1}, and CH, for Galactic sight lines in the \ion{C}{1} survey of JT11, indicate that the column density of \ion{C}{1}$_{\rm tot}$ is well-correlated ($r$ = 0.91--0.98) with the column densities of \ion{Na}{1}, \ion{K}{1}, and CH. 
The slopes of those relationships, 0.85--0.90 in log-log plots, are smaller than those found previously, due to our choice of \ion{C}{1} $f$-values.
$N$(\ion{C}{1}$_{\rm tot}$) is somewhat less well correlated ($r$ = 0.72--0.81) with the column densities of H$_2$ and total hydrogen H$_{\rm tot}$, likely due to variations in local environmental conditions (e.g., the ambient UV radiation field), which affect the abundances of the trace neutral species.
In the LMC and SMC, $N$(\ion{C}{1}$_{\rm tot}$) is consistent with the Galactic trends versus $N$(\ion{Na}{1}) and $N$(CH), but is slightly lower versus $N$(\ion{K}{1}) and $N$(H$_2$).
Like $N$(\ion{Na}{1}) and $N$(\ion{K}{1}), $N$(\ion{C}{1}$_{\rm tot}$) is generally significantly lower, for a given $N$(H$_{\rm tot}$), in the LMC and (especially) in the SMC -- likely due to the combined effects of the lower metallicities, generally stronger radiation fields, and less grain-assisted recombination there.
Such metallicity-related effects may account for a significant part of the scatter in plots of $N$(\ion{C}{1}$_{\rm tot}$) versus $N$(H$_2$) and $N$(H$_{\rm tot}$) for quasar absorption-line systems.

For individual LMC or SMC components with well determined column densities for all three \ion{C}{1} fine-structure states, the implied thermal pressures are typically at least several times higher than those found for most sight lines in the local Galactic ISM, and, in general, the fractions of gas at very high pressures also are higher.
The highest thermal pressures are found for complex regions exhibiting increased activity (e.g., from supernova remnants, star formation, and/or stellar winds); conversely, the lowest thermal pressures are found for simpler, more isolated and/or quiescent regions.
Even those lower thermal pressures are generally higher than typical Galactic values, however.
The higher thermal pressures in the LMC and SMC are consistent with predictions from models of predominantly neutral clouds in systems with lower metallicities, lower dust-to-gas ratios, and stronger UV radiation fields -- where higher thermal pressures are needed to obtain stable cold, neutral clouds.
The similarly high \ion{C}{1} excitation seen for a number of QSO absorption-line systems (often with even lower metallicities) likely reflects varying combinations of those metallicity-related effects, local environmental conditions, and the higher temperature of the CMB at higher $z$.
In general, analysis of the \ion{C}{1} fine-structure excitation in quasar absorbers thus can yield only upper limits on $T_{\rm CMB}$($z$).

\acknowledgments

We acknowledge helpful comments on the initially submitted version of the paper by S. Federman, E. Jenkins, and the referee (J. van Loon).
Support for {\it HST} guest observer programs 8145, 9757, and 12978 and for archival program 10692 was provided by NASA through grants from the Space Telescope Science Institute, which is operated by the Association of Universities for Research in Astronomy, Inc., under NASA contract NAS 5-26555.

{\it Facilities:} \facility{HST (STIS)}, \facility{ESO: 3.6m (CES, HARPS)}. \facility{VLT: Kueyen (UVES)}, \facility{Max Planck: 2.2m (FEROS)}

\clearpage

\appendix

\section{\ion{C}{1} Spectra}
\label{sec-apc1spec}

\begin{figure}
\epsscale{0.8}
\plotone{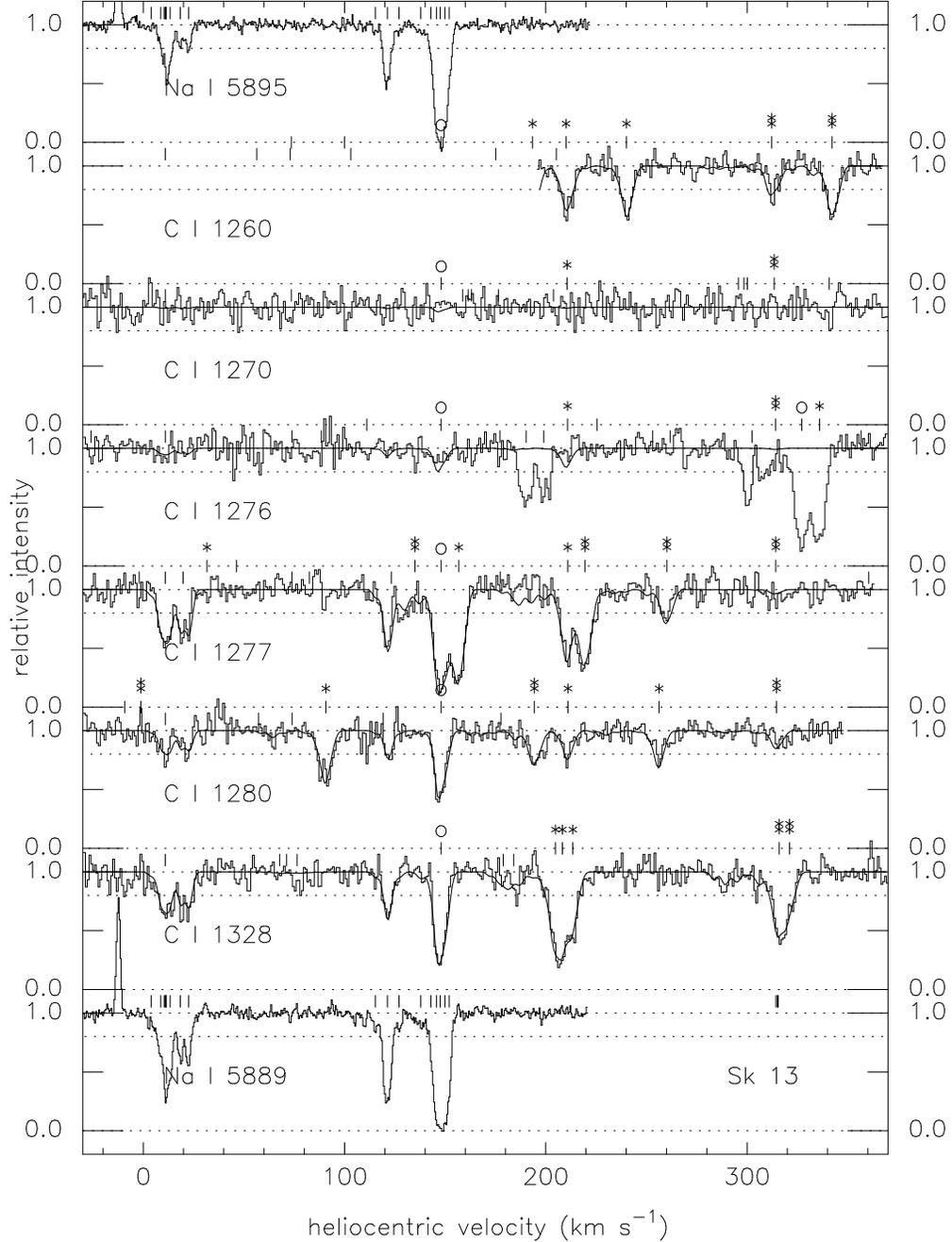}
\caption{{\it HST}/STIS E140H spectra (FWHM = 2.7 km~s$^{-1}$) of \ion{C}{1} multiplets toward the SMC star Sk 13 (with \ion{Na}{1} to show the detailed component structure).
Histograms give the normalized observed spectra; smooth curves give the theoretical profiles for the adopted component structure (Table~\ref{tab:comps}).
Absorption from the \ion{C}{1} ground state at $v$ $<$ 80 km~s$^{-1}$ is from the Galactic disk and halo; absorption at $v$ $>$ 80 km~s$^{-1}$ is from gas in the SMC.
For each \ion{C}{1} spectrum, the lower and upper tick marks indicate the locations of the main Galactic and SMC absorption components, respectively; most of the absorption is from \ion{C}{1}.
For the SMC components, the \ion{C}{1} ground and excited fine-structure states are noted by o, *, or $\stackrel{\textstyle *}{*}$.}
\label{fig:sk13c1}
\end{figure}

\begin{figure}
\epsscale{0.8}
\plotone{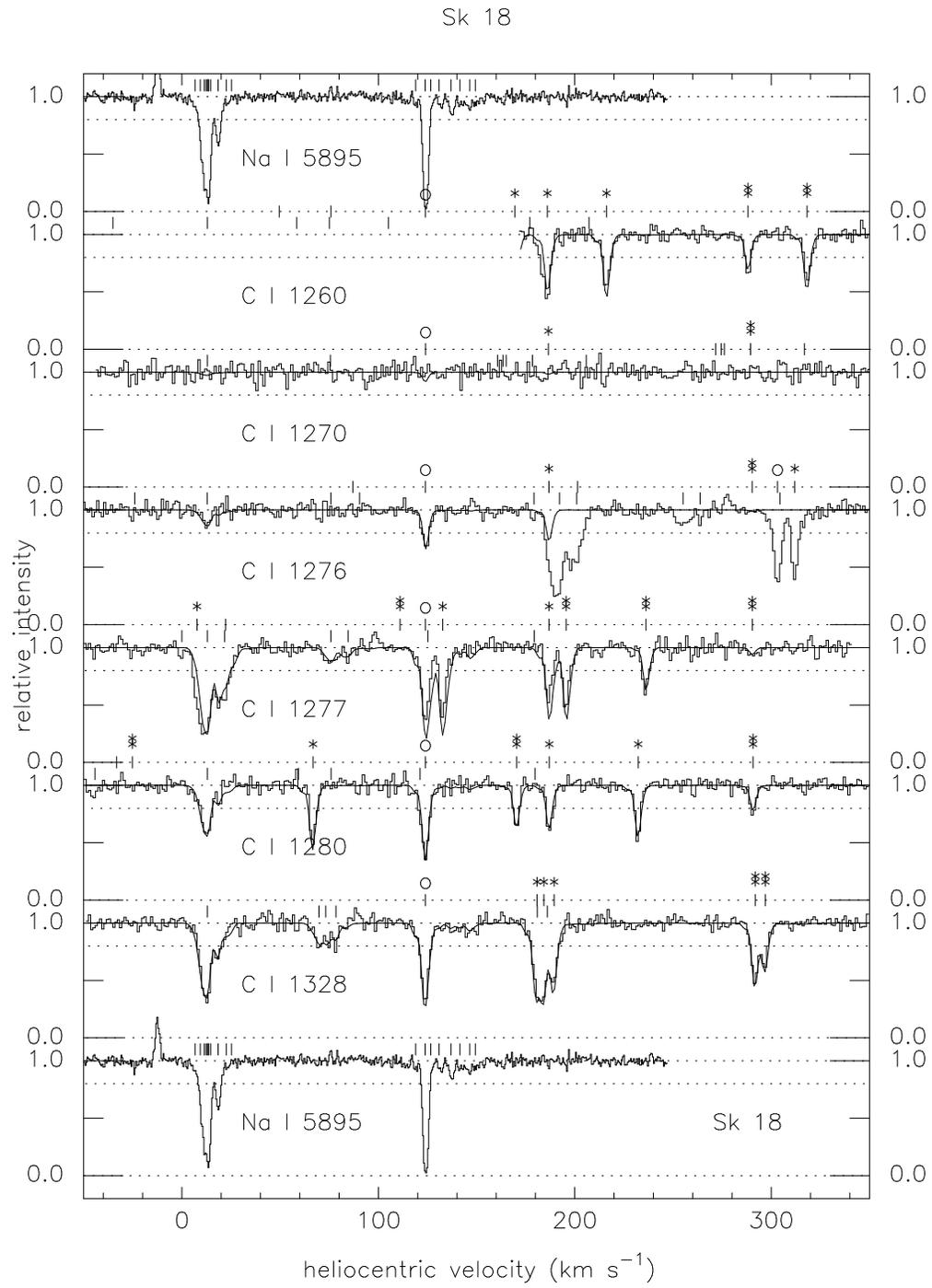}
\caption{Same as Fig.~\ref{fig:sk13c1}, but for the SMC star Sk 18.}
\label{fig:sk18c1}
\end{figure}

\begin{figure}
\epsscale{0.8}
\plotone{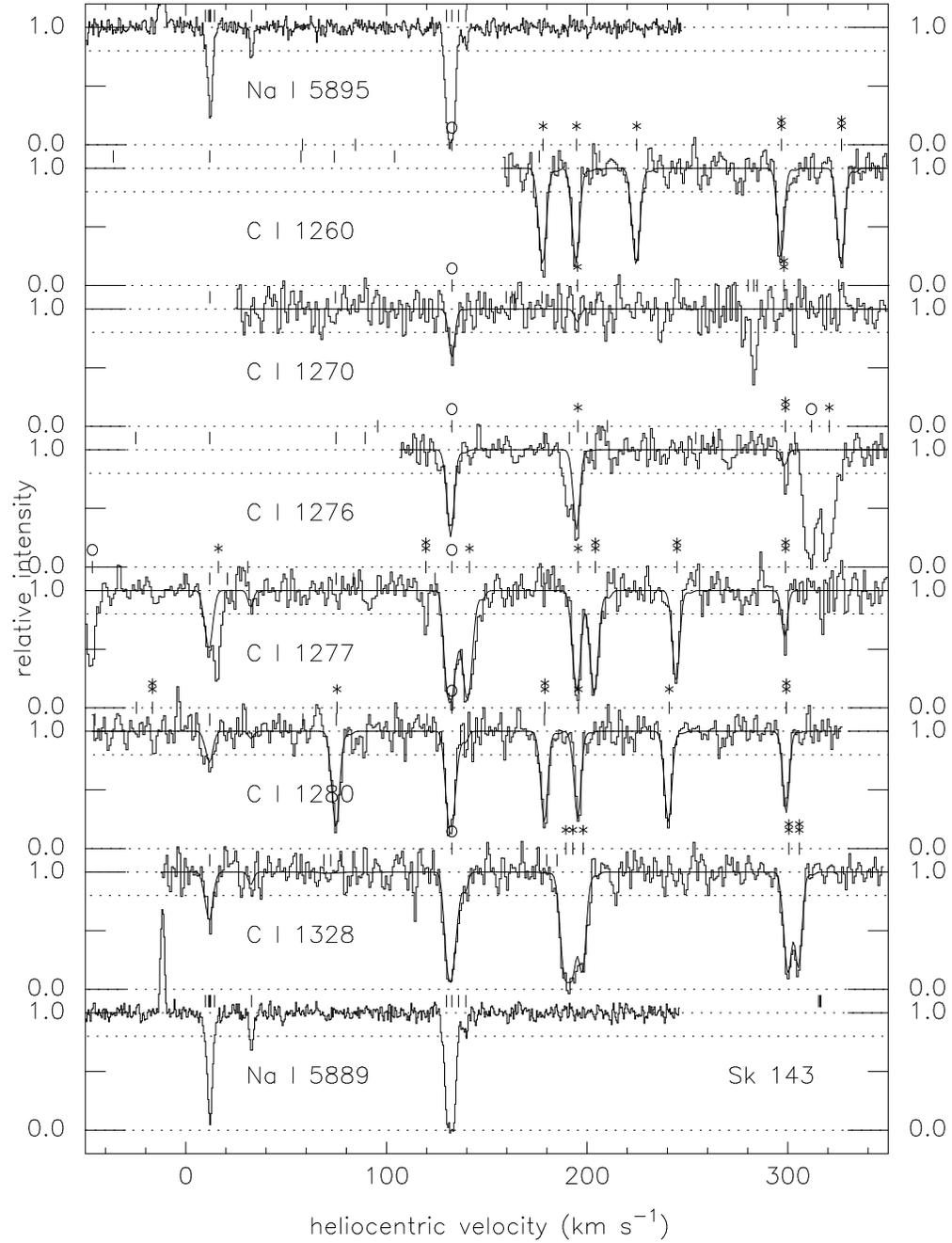}
\caption{Same as Fig.~\ref{fig:sk13c1}, but for the SMC star Sk 143.
Absorption near 285 km~s$^{-1}$ for \ion{C}{1} $\lambda$1270 is due to \ion{S}{1}.}
\label{fig:sk143c1}
\end{figure}

\begin{figure}
\epsscale{0.8}
\plotone{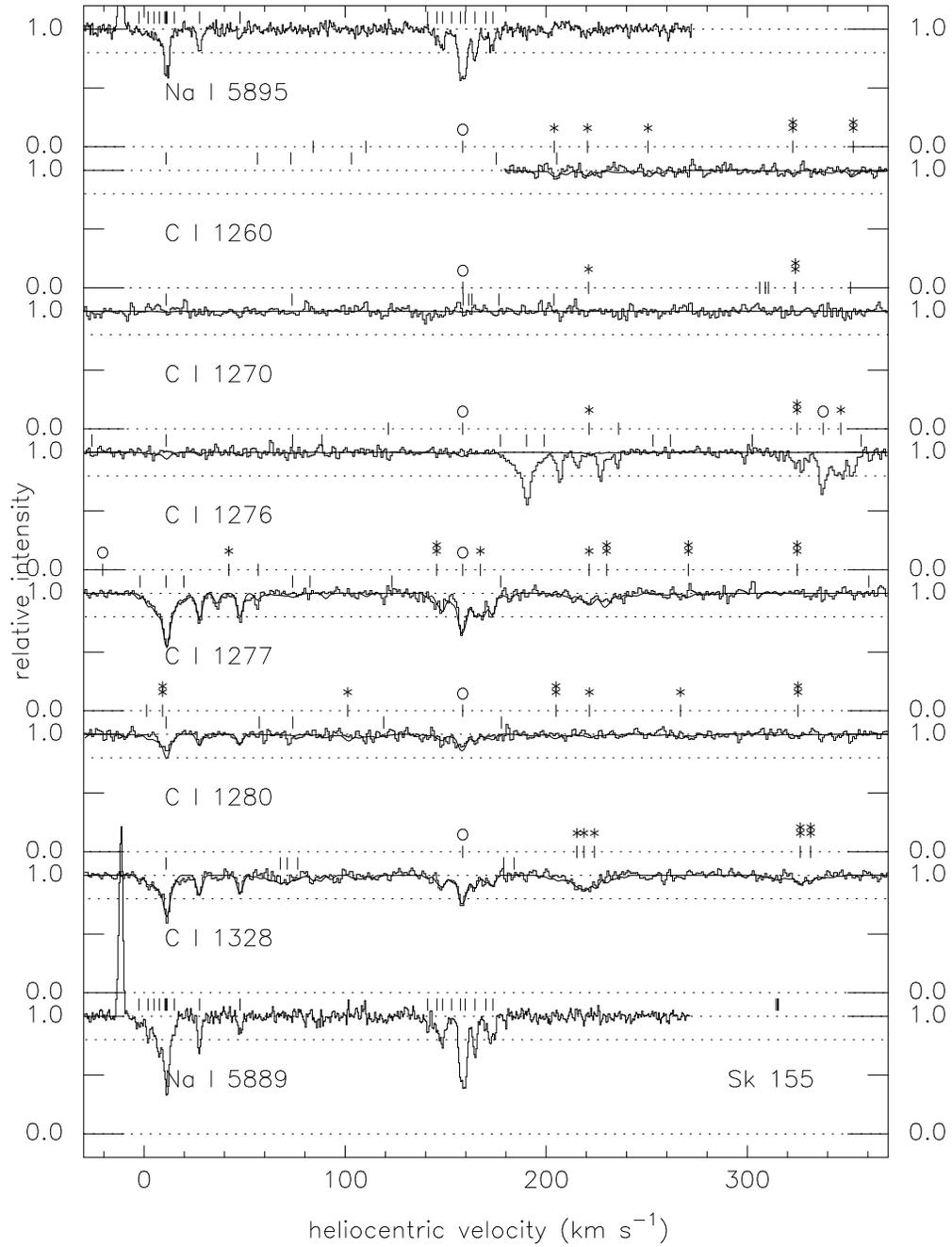}
\caption{Same as Fig.~\ref{fig:sk13c1}, but for the SMC star Sk 155.}
\label{fig:sk155c1}
\end{figure}

\begin{figure}
\epsscale{0.8}
\plotone{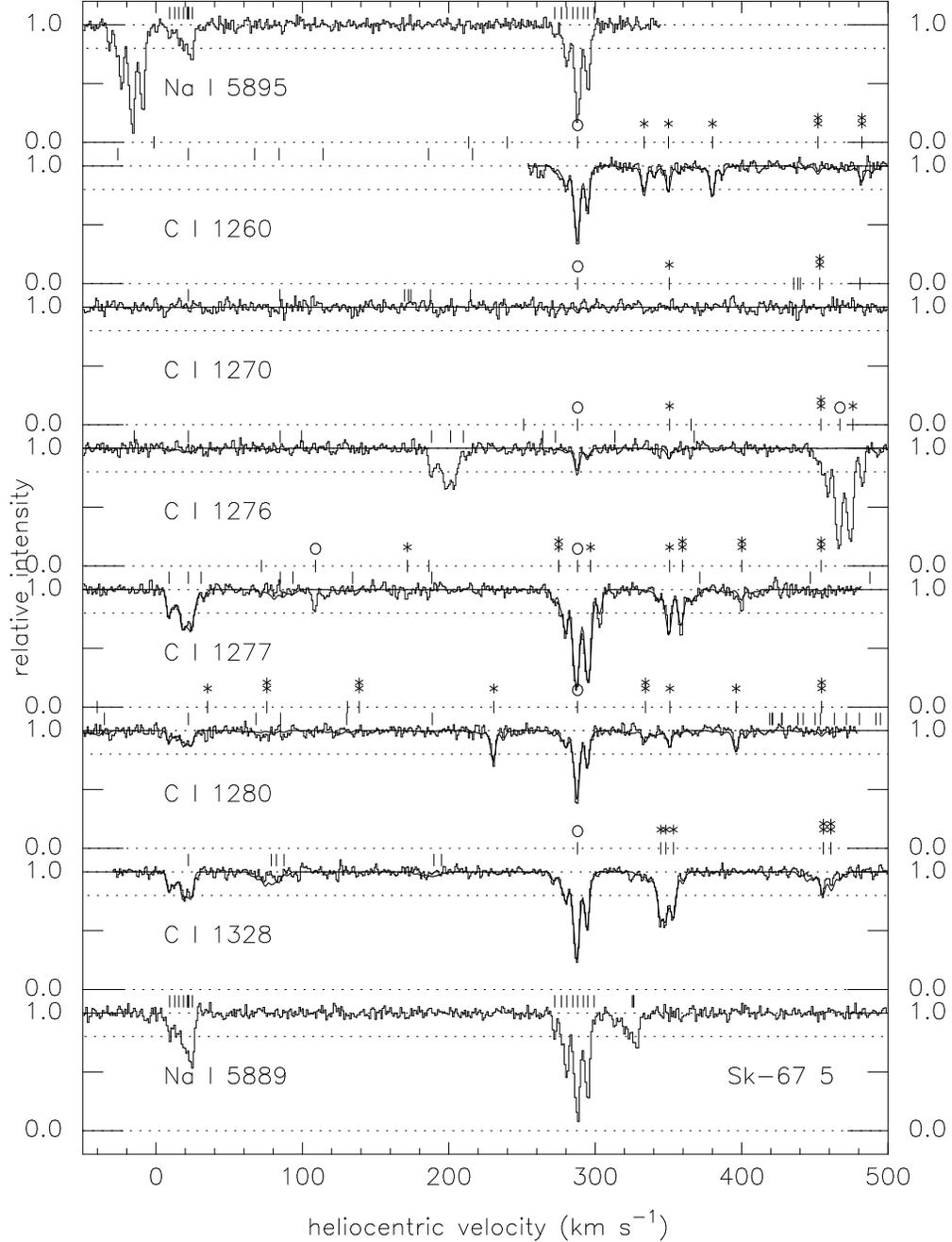}
\caption{{\it HST}/STIS E140H spectra (FWHM = 2.7 km~s$^{-1}$) of \ion{C}{1} multiplets toward the LMC star Sk$-$67~5 (with \ion{Na}{1} to show the detailed component structure).
Histograms give the normalized observed spectra; smooth curves give the theoretical profiles for the adopted component structure (Table~\ref{tab:comps}).
Absorption from the \ion{C}{1} ground state at $v$ $<$ 100 km~s$^{-1}$ is from the Galactic disk and halo; absorption at $v$ $>$ 100 km~s$^{-1}$ is from gas in the LMC.
For each \ion{C}{1} spectrum, the lower and upper tick marks indicate the locations of the main Galactic and LMC absorption components, respectively; most of the absorption is from \ion{C}{1}.
For the LMC components, the \ion{C}{1} ground and excited fine-structure states are noted by o, *, or $\stackrel{\textstyle *}{*}$.}
\label{fig:sk67d5c1}
\end{figure}

\begin{figure}
\epsscale{0.8}
\plotone{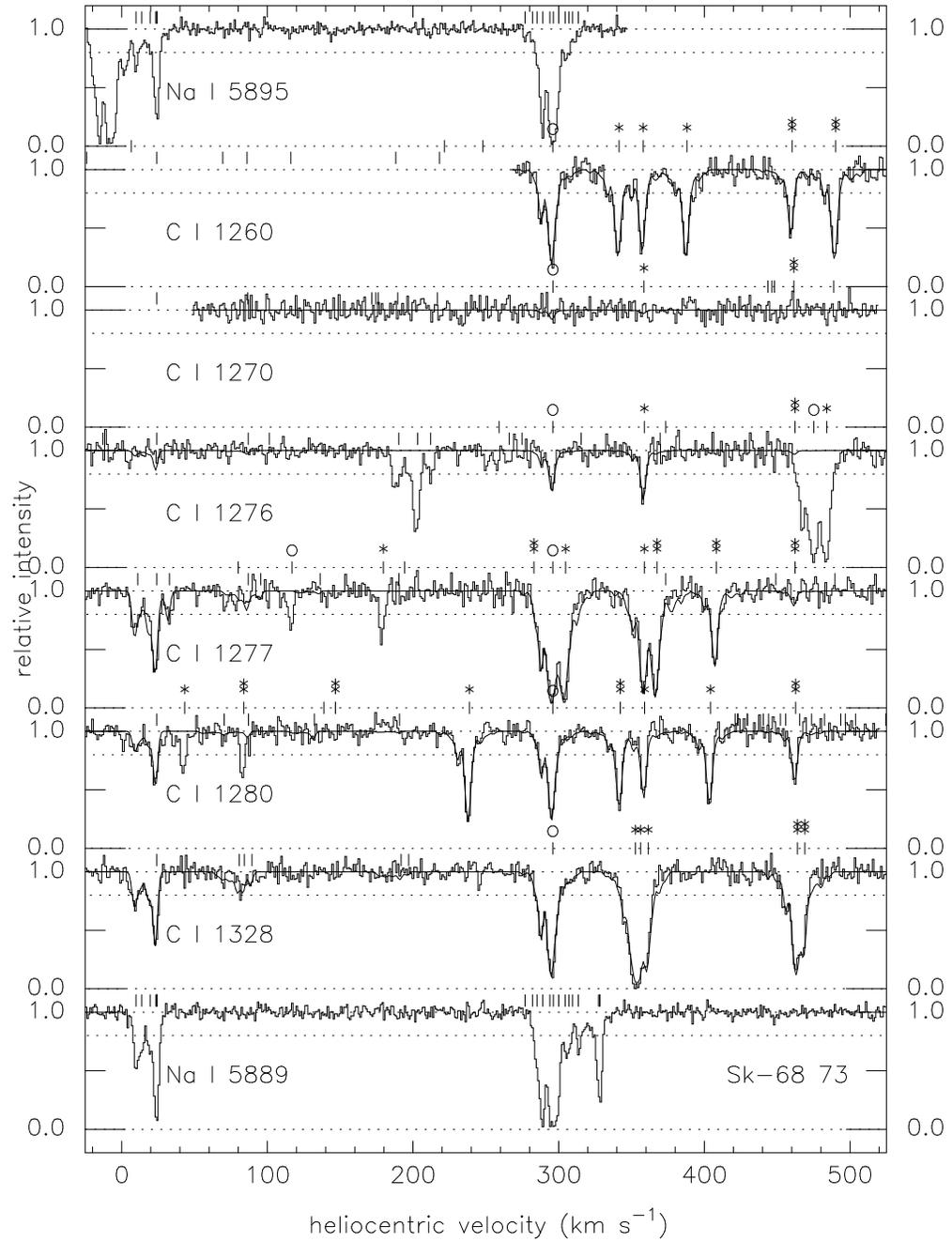}
\caption{Same as Fig.~\ref{fig:sk67d5c1}, but for the LMC star Sk$-$68~73.}
\label{fig:sk68d73c1}
\end{figure}

\begin{figure}
\epsscale{0.8}
\plotone{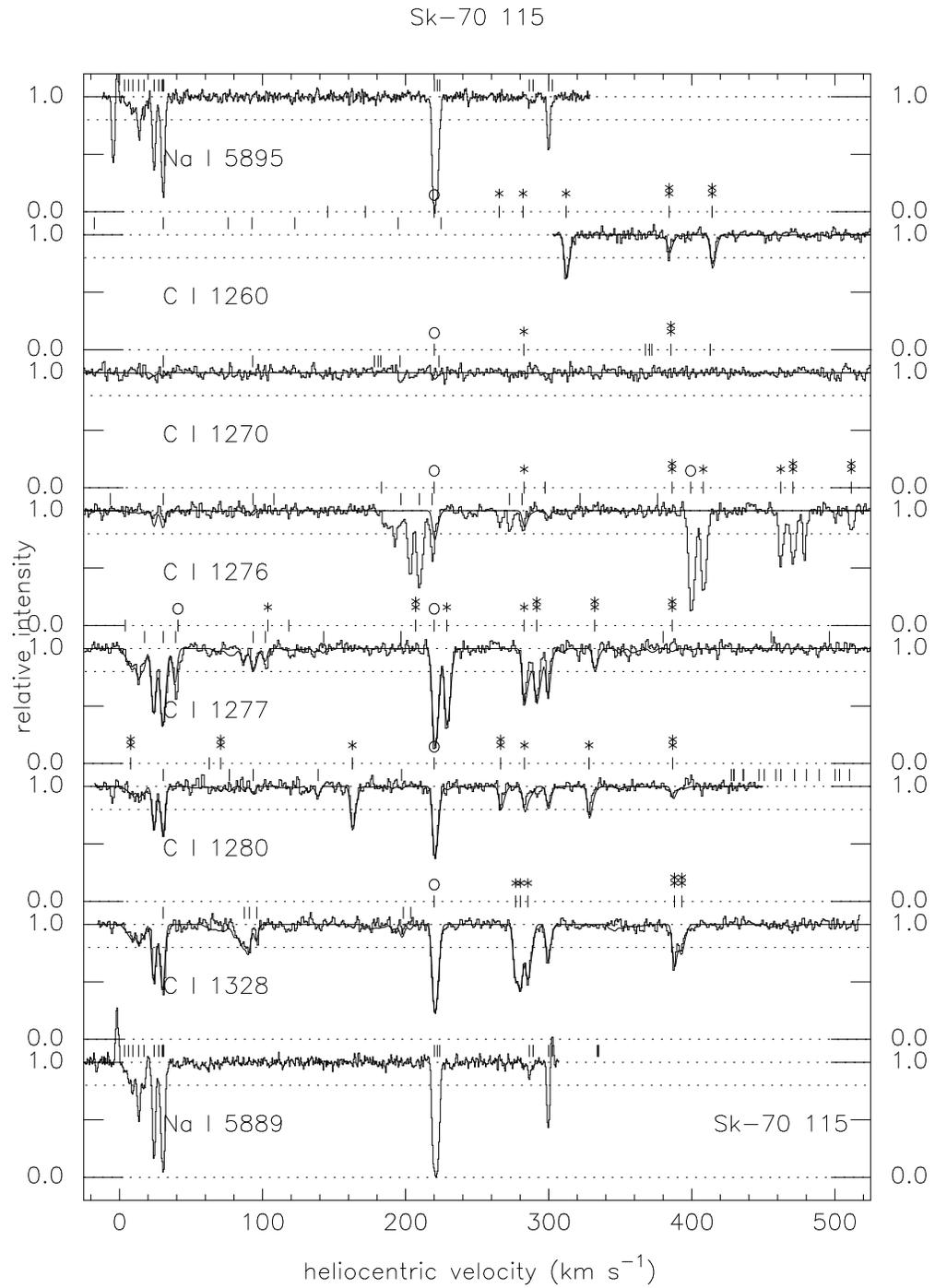}
\caption{Same as Fig.~\ref{fig:sk67d5c1}, but for the LMC star Sk$-$70~115.}
\label{fig:sk70d115c1}
\end{figure}

Figures~\ref{fig:sk13c1} through \ref{fig:sk70d115c1} show the STIS spectra of all six of the \ion{C}{1} multiplets considered in this study (Table~\ref{tab:c1}), toward the seven Magellanic Clouds stars in our sample.
The corresponding higher resolution spectra of the \ion{Na}{1} $\lambda$5895 and $\lambda$5889 lines are included at the top and bottom, respectively, to show the detailed component structure derived for each sight line.
The histograms give the normalized observed spectra, while the smooth curves give the theoretical profiles computed using the adopted component structures listed in Table~\ref{tab:comps} (in each case for the lines of the given multiplet only), smoothed to the STIS echelle resolution (FWHM $\sim$ 2.7 km~s$^{-1}$).
The tick marks above the \ion{Na}{1} profiles denote the detailed component structure; the lower and upper tick marks above the \ion{C}{1} profiles denote the locations of the strongest Galactic and Magellanic Clouds components, respectively, for the ground and excited state lines of each multiplet (marked by o, *, and $\stackrel{\textstyle *}{*}$ for the main SMC or LMC component).
Absorption from the \ion{C}{1} ground state at $v$ $<$ 80 km~s$^{-1}$ is from the Galactic disk and halo; absorption from gas at $v$ $>$ 80 km~s$^{-1}$ is from gas in the SMC; the break point is assumed to be 100 km~s$^{-1}$ for the LMC sight lines.

\section{Component Structures for \ion{Na}{1} and \ion{C}{1}}
\label{sec-apcomp}

Table~\ref{tab:comps} lists the detailed component structures ($N$, $b$, $v$) derived from fits to the high-resolution CES spectra (FWHM $\sim$ 1.35--2.0 km~s$^{-1}$) of the \ion{Na}{1} $\lambda$5895 and $\lambda$5889 lines, with the $N$ further constrained by fits to the corresponding lower-resolution UVES spectra (FWHM $\sim$ 4.5--4.9 km~s$^{-1}$) of both those lines and the weaker $\lambda$3302 doublet (see Welty et al. 2006 and Welty \& Crowther, in preparation).
The $b$-values of the strongest \ion{Na}{1} components were adjusted to yield consistent fits to both the weak and the strong \ion{Na}{1} doublets.
The $b$-values and the relative $v$ were then fixed for fits to the STIS spectra (FWHM $\sim$ 2.7 km~s$^{-1}$) of the six \ion{C}{1} multiplets, with the $b$-values of the strongest components increased slightly to yield consistent fits to both the weakest and strongest multiplets.

\section{Galactic Column Densities}
\label{sec-apgal}

Comparisons of the column density of \ion{C}{1}$_{\rm tot}$ with the corresponding column densities of other interstellar species can provide insights into both the behavior of \ion{C}{1} and local physical conditions in the ISM (e.g., Welty \& Hobbs 2001).
Unfortunately, data for the usual optical tracers of interstellar material (\ion{Na}{1}, \ion{K}{1}, \ion{Ca}{1}, \ion{Ca}{2}, \ion{Ti}{2}, CH, CH$^+$, CN) have not been available (or else have not been reported) for many of the sight lines analyzed by JT11.
Moreover, the UV spectra of a number of more commonly observed stars with previously reported $N$(\ion{C}{1}$_{\rm tot}$) (e.g., Jenkins \& Shaya 1979; Jenkins et al. 1983; Wolfire et al. 2008) -- which often do have data for the optical tracers -- have not been re-analyzed using the $f$-values adopted by JT11.
We have therefore sought to augment the existing column density data for other species in the JT11 sight line sample by analyzing any existing optical spectra (from the ESO public archive and/or from an archive of spectra obtained for studies of the diffuse interstellar bands\footnotemark) and by obtaining new optical spectra with Magellan/MIKE.
\footnotetext{The ESO public archive (at archive.eso.org) contains spectra from FEROS, HARPS, and UVES.  
The DIB database (at dib.uchicago.edu/public) contains spectra obtained with the ARC echelle spectrograph (e.g., Friedman et al. 2011).}
Because high-resolution spectra suitable for detailed profile fits are not available in most cases, the column densities generally were obtained from the equivalent widths and/or AOD integrations of relatively weak lines in the spectra (e.g., $\la$ 15 m\AA\ for CH).
For \ion{Na}{1}, all the column density values greater than 6 $\times$ 10$^{13}$ cm$^{-2}$ were derived from observations of the weaker $\lambda$3302 doublet (''U'' lines) -- so that saturation should not be an issue for those highest values.
Lower limits to the column densities are given where no sufficiently weak lines are available.
We have also derived some new atomic hydrogen column densities from the (damped) Lyman-$\alpha$ absorption seen in the same UV spectra analyzed by JT11, using the usual continuum reconstruction method.
The currently available column density data for H, H$_2$, \ion{Na}{1}, \ion{K}{1}, and CH for the JT11 sight line sample are listed in Table~\ref{tab:gal}.

\section{Carbon Reference Abundances}
\label{sec-apcref}

Table~\ref{tab:crefab} lists values of the carbon abundance derived from different kinds of objects in the LMC and SMC, obtained from the literature.
Estimates of the Solar carbon abundance appear to have stabilized at a value of about 8.40 dex, on the usual logarithmic scale in which the hydrogen abundance is 12.0 (Lodders 2003; Asplund et al. 2009), and most of the LMC and SMC values for each object type also are in reasonable agreement, given the differences in samples and methodology and the typical uncertainties.
The values derived from later-type giants and supergiants may be slightly higher than those obtained from \ion{H}{2} regions and earlier-type stars, however.
Our adopted values -- 7.95 dex for the LMC and 7.45 dex for the SMC -- are consistent with several other recent estimates also based on values compiled from studies of different kinds of objects (e.g., Hunter et al. 2007; Salmon et al. 2012; Tchernyshyov et al. 2015).
Those adopted values are about 0.2 dex lower, relative to Solar abundances, than the values obtained for most heavier elements -- which are typically about $-$0.3 dex for the LMC and about $-$0.7 dex for the SMC (largely from the same references).
While the depletion of carbon in the ISM of the Magellanic Clouds is currently unknown, it is unlikely to be more severe than the roughly $-$0.2 dex found for the local Galactic ISM (e.g., Jenkins 2009).

\section{\ion{C}{1} in QSO Absorption-line Systems}
\label{sec-apqso}

Table~\ref{tab:c1qso} lists the column densities of \ion{C}{1} in QSO absorption-line systems that are used in Figures~\ref{fig:c1qso} and \ref{fig:fsqso}.
The top section of the table lists systems for which all three fine-structure states have been detected (Quast et al. 2002; Srianand et al. 2005, 2008; Noterdaeme et al. 2007, 2010, 2015; Jorgenson et al. 2010; Carswell et al. 2011; Guimar\~{a}es et al. 2012; Ma et al. 2015). 
The bottom section lists systems for which only total \ion{C}{1} column densities are available; in some cases, the values are for just the ground fine-structure state.
In most of the upper systems, the values are for ''individual'' components determined in fits to the absorption-line profiles seen in moderately high resolution spectra (FWHM generally $>$ 5 km~s$^{-1}$) obtained with Keck/HIRES and/or VLT/UVES.
Those references used $f$-values from Morton (2003) (or similar values from earlier work), so that the absolute column densities are not directly comparable to those in the JT11 sample.
Many of the values were derived primarily from observations of the stronger \ion{C}{1} multiplets, however, for which the differences in $f$ are relatively small (less than a factor of 2).
In many cases, the inferred \ion{C}{1} column densities are not extremely high, so the relative level populations $f_1$ and $f_2$ may be reasonably well determined.
We note, however, that some of the $b$-values are either very low or very high, in comparison with those derived from higher resolution spectra of \ion{C}{1} and other trace neutral species in our Galaxy and in the Magellanic Clouds -- which will affect the derived column densities.
(And, for cases with very low $b$-values, use of the JT11 $f$-values would yield both somewhat larger $b$ and lower column densities.)
Some of the points are located (nominally) outside the ''allowed'' region bounded on the left by the line between the minimum ($f_1$,$f_2$) [set by $T$(CMB)] and the ''infinite density'' point [(0.33,0.56), set by the degeneracies of the fine-structure levels], and on the right by the range of theoretical curves for ''reasonable'' $T$.
The column densities of H, H$_2$, and the first \ion{C}{1}$_{\rm tot}$ refer to the entire system, and the derived quantities $f$(H$_2$) and [Zn/H] are average values which may not reflect the actual conditions in some of the individual components. 
For systems where $N$(\ion{Zn}{2}) has not been measured, the [Zn/H] values (in parentheses) are metallicities estimated from \ion{O}{1}, \ion{Si}{2}, and/or \ion{S}{2} (from the original references).
Systems in the bottom section of the table taken from Srianand et al. (2005) generally do not exhibit detectable H$_2$ absorption; most of those systems also are characterized by metallicities less than 0.05 $\times$ Solar.
The two columns following the fractional populations ($f_1$ and $f_2$) give the excitation temperatures for \ion{C}{1} $N$(0)/$N$(1) -- as expected from the Cosmic Microwave Background at the redshift of the absorber and as derived from the fitted column densities.

\clearpage

\clearpage

\begin{deluxetable}{lcccrrrrr}
\tablecolumns{9}
\tabletypesize{\scriptsize}
\tablecaption{SMC and LMC Component Structures \label{tab:comps}}
\tablewidth{0pt}

\tablehead{
\multicolumn{1}{c}{Star}&
\multicolumn{1}{c}{Group}&
\multicolumn{1}{c}{$v$}&
\multicolumn{1}{c}{$b$\tablenotemark{a}}&
\multicolumn{1}{c}{$N$(Na~I)}&
\multicolumn{1}{c}{$N$(C~I$^0$)}&
\multicolumn{1}{c}{$N$(C~I$^*$)}&
\multicolumn{1}{c}{$N$(C~I$^{**}$)}&
\multicolumn{1}{c}{C~I$_{\rm tot}$/Na~I}\\
\multicolumn{2}{c}{ }&
\multicolumn{1}{c}{(km~s$^{-1}$)}&
\multicolumn{1}{c}{(km~s$^{-1}$)}&
\multicolumn{1}{c}{(10$^{11}$ cm$^{-2}$)}&
\multicolumn{1}{c}{(10$^{12}$ cm$^{-2}$)}&
\multicolumn{1}{c}{(10$^{12}$ cm$^{-2}$)}&
\multicolumn{1}{c}{(10$^{12}$ cm$^{-2}$)}}

\startdata
Sk 13       &   &   4.0 & [1.5]       &  0.4 & $<$1.0 &        &        & \\
            &   &   8.7 & [1.5]       &  1.6 &    4.3 &        &        & \\
            &   &  11.3 & [0.4]       &  2.8 &    2.6 &        &        & \\
            &   &  13.4 &  1.3        &  2.3 &    3.5 &        &        & \\
            &   &  18.4 &  1.4        &  1.7 &    4.2 &        &        & \\
            &   &  22.5 &  1.5        &  2.0 &    4.8 &        &        & \\
            & A & 115.3 & [1.0]       &  0.3 & $<$1.2 & $<$1.8 & $<$2.1 & \\
            & A & 121.3 &  2.1        &  6.2 &    8.3 &    3.1 &    1.9 &   21.4 \\
\enddata
\tablecomments{Velocities are heliocentric.  All column densities are $\ge$ 2$\sigma$; limits are 3$\sigma$.}
\tablenotetext{a}{Where two values are given, the first is for \ion{Na}{1} and the second is for \ion{C}{1}.}
\end{deluxetable}

\clearpage

\begin{deluxetable}{lrrrrrrrrrrrrr}
\tablecolumns{14}
\tabletypesize{\scriptsize}
\tablecaption{Galactic Column Densities: H, H$_2$, \ion{C}{1}$_{\rm tot}$, \ion{Na}{1}, \ion{K}{1}, CH \label{tab:gal}}
\tablewidth{0pt}

\tablehead{
\multicolumn{1}{c}{HD}&
\multicolumn{1}{c}{Name}&
\multicolumn{1}{c}{$N$(H)}&
\multicolumn{1}{c}{Ref}&
\multicolumn{1}{c}{$N$(H$_2$)}&
\multicolumn{1}{c}{Ref}&
\multicolumn{1}{c}{$N$(C~I$_{\rm tot}$)}&
\multicolumn{1}{c}{Ref}&
\multicolumn{1}{c}{$N$(Na~I)}&
\multicolumn{1}{c}{Ref}&
\multicolumn{1}{c}{$N$(K~I)}&
\multicolumn{1}{c}{Ref}&
\multicolumn{1}{c}{$N$(CH)}&
\multicolumn{1}{c}{Ref}}

\startdata
HD   108   &        &    21.36 &  1 &  \nodata &    & $>$14.98 &  1 &  \nodata &    &  \nodata &    &  \nodata &    \\
HD  1383   &        &    21.42 &  2 &    20.45 &  1 & $>$14.80 &  1 &  \nodata &    &  \nodata &    &  \nodata &    \\
HD  3827   &        &    20.56 &  1 &  \nodata &    &    13.57 &  1 &  \nodata &    &  \nodata &    &  \nodata &    \\
HD 15137   &        &    21.11 &  3 &    20.27 &  2 &    14.62 &  1 &  \nodata &    &  \nodata &    &    12.79 &  1 \\
HD 23478   &        &    20.71 &  4 &    20.48 &  1 & $>$14.75 &  1 &  \nodata &    &  \nodata &    &    13.34 &  2 \\
HD 24190   &        &    21.10 &  2 &    20.38 &  1 &    14.45 &  1 &  \nodata &    &  \nodata &    &    12.98 &  2 \\
HD 24534   & X Per  &    20.73 &  3 &    20.92 &  3 &    14.98 &  2 &    14.18 &  1 &    12.09 &  1 &    13.55 &  1 \\
HD 27778   & 62 Tau &    20.95 &  1 &    20.79 &  3 &    15.06 &  2 &    13.95 &  1 &    11.84 &  1 &    13.48 &  1 \\
\enddata
\tablecomments{Column densities (cm$^{-2}$) are logarithmic.
For Figs.~\ref{fig:c1abund} and \ref{fig:c1qso}, the Trapezium region sight lines are HD 37021 and HD 37061; the Sco-Oph region sight line is HD 147888.}
\tablerefs{
H: 1 = this paper; 2 = Cartledge et al. 2004, 2008; 3 = Diplas \& Savage 1994; 4 = Jenkins 2009; 5 = Jensen \& Snow 2007; 6 = Welty et al., in preparation; 7 = Andr\'{e} et al. 2003; 8 = Rachford et al. 2002 \\
H$_2$: 1 = Cartledge et al. 2004, 2008; 2 = Sheffer et al. 2008; 3 = Rachford et al. 2002, 2009; 4 = Jensen \& Snow 2007; 5 = Andr\'{e} et al. 2003; 6 = Snow et al. 2007, 2008; 7 = Burgh et al. 2007; 8 = Pan et al. 2005 \\
C~I: 1 = JT11; 2 = Burgh et al. 2010; 3 = Jenkins 2002; 4 = Welty et al., in preparation; 5 = Meyer et al. 2012: 6 = this paper \\
Na~I: 1 = Welty (U lines); 2 = Welty (D lines); 3 = Sembach et al. 1993 (D); 4 = Meyer et al. 2012; 5 = Hunter et al. 2006 (U); 6 = this paper (D) \\
K~I: 1 = Welty; 2 = Pan et al. 2004 \\
CH: 1 = Welty; 2 = Sheffer et al. 2008; 3 = Pan et al. 2005}
\end{deluxetable}

\begin{deluxetable}{lrrrrrrl}
\tablecolumns{8}
\tabletypesize{\scriptsize}
\tablecaption{Carbon reference abundances \label{tab:crefab}}
\tablewidth{0pt}

\tablehead{
\multicolumn{1}{c}{Objects}&
\multicolumn{1}{c}{Milky Way}&
\multicolumn{1}{c}{N}&
\multicolumn{1}{c}{LMC}&
\multicolumn{1}{c}{N}&
\multicolumn{1}{c}{SMC}&
\multicolumn{1}{c}{N}&
\multicolumn{1}{c}{Reference}}

\startdata
Solar photosphere     & 8.39 & 1 &      &   &      &   & Lodders 2003 \\
                      & 8.43 & 1 &      &   &      &   & Asplund et al. 2009 \\
\hline
H~II regions          &      &   & 7.93 & 4 & 7.28 & 3 & Dufour 1991 \\
                      &      &   & 7.81 & 4 & 7.20 & 3 & Kurt \& Dufour 1998 \\
                      &      &   & 7.90 & 2 & 7.40 & 1 & Garnett 1999 \\
                      &      &   &      &   & 7.53 & 1 & Kurt et al. 1999 \\
                      &      &   & 8.03 & 6 & 7.39 & 5 & Reyes 1999 \\
\hline
O--B main sequence    &      &   &      &   & 6.80 & 3 & Rolleston et al. 1993 \\
\enddata
\tablecomments{Abundances are logarithmic, on a scale where the abundance of hydrogen is 12.00.}
\end{deluxetable}

\renewcommand{\tabcolsep}{5.0pt}
\begin{deluxetable}{lcrcrccrcccrrr}
\tablecolumns{14}
\tabletypesize{\scriptsize}
\tablecaption{\ion{C}{1} in QSO Absorption-line Systems \label{tab:c1qso}}
\tablewidth{0pt}

\tablehead{
\multicolumn{1}{c}{QSO}&
\multicolumn{1}{c}{$N$(H)}&
\multicolumn{1}{c}{$N$(H$_2$)}&
\multicolumn{1}{c}{$f$(H$_2$)}&
\multicolumn{1}{c}{$N$(C~I$_{\rm tot}$)}&
\multicolumn{1}{c}{[Zn/H]}&
\multicolumn{1}{c}{$z_{\rm abs}$}&
\multicolumn{1}{c}{$b$}&
\multicolumn{1}{c}{$N$(C I$_{\rm tot}$)}&
\multicolumn{1}{c}{$f_1$}&
\multicolumn{1}{c}{$f_2$}&
\multicolumn{1}{c}{$T_{\rm CMB}$}&
\multicolumn{1}{c}{$T_{01}$}&
\multicolumn{1}{c}{Ref}}

\startdata
0013$-$004&20.83 & 18.86 & 0.0210 &  14.27 & $-$0.73 & 1.96822 &  4.3 & 13.47 & 0.43$\pm$0.01 & 0.20$\pm$0.01 &  8.1 & 24.7 & 1 \\ 
         &       &       &        &        &         & 1.97296 &  6.2 & 13.67 & 0.34$\pm$0.01 & 0.07$\pm$0.01 &      & 14.4 &   \\ 
0154+194 & 21.75 &\nodata& \nodata&  13.95 & $-$0.72 & 2.25152 &  5.2 & 13.77 & 0.50$\pm$0.08 & 0.09$\pm$0.02 &  8.9 & 26.0 & 2 \\
         &       &       &        &        &         & 2.25163 &  6.3 & 13.49 & 0.51$\pm$0.21 & 0.12$\pm$0.07 &      & 29.7 &   \\
0515$-$414&19.88 & 16.94 & 0.0023 &  13.93 & $-$0.40 & 1.15079 &  2.0 & 13.79 & 0.36$\pm$0.01 & 0.09$\pm$0.01 &  5.9 & 15.7 & 3 \\ 
         &       &       &        &        &         & 1.15085 &  3.5 & 13.37 & 0.25$\pm$0.01 & 0.05$\pm$0.01 &      & 11.1 &   \\
0551$-$366&20.70 & 17.42 & 0.0010 &  13.78 & $-$0.31 & 1.96214 &  2.1 & 13.03 & 0.46$\pm$0.15 & 0.12$\pm$0.11 &  8.1 & 23.0 & 1 \\ 
         &       &       &        &        &         & 1.96221 & 12.8 & 13.41 & 0.37$\pm$0.09 & 0.07$\pm$0.07 &      & 15.7 &   \\
\enddata
\tablecomments{Column densities (cm$^{-2}$) are logarithmic.  Limits are 3 $\sigma$.
The values for $N$(H), $N$(H$_2$), $f$(H$_2$), the first $N$(C~I$_{\rm tot}$), and [Zn/H] are for the entire system; the values for the second $N$(C~I$_{\rm tot}$), $f_1$, and $f_2$ are for the components (or component groups) at redshift $z_{\rm abs}$ within each system.
Systems in the lower part of the table do not have complete information on the excitation of \ion{C}{1} -- so only the total column density is given.
Metallicities given in parentheses were estimated from the column densities of \ion{O}{1}, \ion{Si}{2}, and/or \ion{S}{2} (in the original references).
The values for $N$(C~I$_{\rm tot}$) in this table were not obtained using the $f$-values from JT11, and so are not directly comparable to the values for Galactic and Magellanic Clouds sight lines given elsewhere in this paper.}
\tablerefs{1 = Srianand et al. 2005; 2 = Noterdaeme et al. 2015; 3 = Quast et al. 2002; 4 = Jorgenson et al. 2010; 5 = Guimar\~{a}es et al. 2012; 6 = Ma et al. 2015; 7 = Noterdaeme et al. 2010; 8 = Carswell et al. 2011; 9 = Srianand et al. 2008; 10 = Noterdaeme et al. 2007a; 11 = Noterdaeme et al. 2007b; 12 = Prochaska \& Wolfe 1996; 13 = Blades et al. 1985; 14 = Srianand et al. 2007; 15 = Dutta et al. 2015; 16 = Boiss\'{e} et al. 1998; 17 = Albornoz V\'{a}squez et al. 2014; 18 = Prochaska et al. 2007; 19 = Ledoux et al. 2006; 20 = Noterdaeme et al. 2009; 21 = Srianand et al. 2014; 22 = Roth \& Bauer 1999; 23 = Muzahid et al. 2016; 24 = Kr\"{u}hler et al. 2013.}
\end{deluxetable}

\end{document}